\documentclass[fleqn,usenatbib]{mnras}

\usepackage[T1]{fontenc}
\usepackage{ae,aecompl}

\usepackage{graphicx}	
\usepackage{amsmath}	
\usepackage{amssymb}	
\usepackage{verbatim}
\usepackage[normalem]{ulem}
\usepackage[dvipsnames]{xcolor}

\usepackage{newtxtext,newtxmath}

\title[MicroJy radio origin in two strongly-lensed RQQ]{Using strong lensing to understand the microJy radio emission in two radio quiet quasars at redshift 1.7}

\author[Hartley et al.]{
P. Hartley,$^{1,2}$\thanks{E-mail: philippahartley@hotmail.com}
N. Jackson,$^{2}$ S. Badole,$^{2}$ J. P. McKean,$^{3,4}$ D. Sluse,$^{5}$ H. Vives-Arias$^{6}$
\\
$^{1}$SKAO, Jodrell Bank, Lower Withington, Macclesfield, Cheshire, SK11 9FT, UK\\
$^{2}$Jodrell Bank Centre for Astrophysics, School of Physics and Astronomy, University of Manchester, Oxford Road, Manchester, M13 9PL, UK\\
$^{3}$ASTRON, Netherlands Institute for Radio Astronomy, Oude Hoogeveensedijk 4, 7991 PD, Dwingeloo, the Netherlands\\
$^{4}$Kapteyn Astronomical Institute, University of Groningen, P.O. Box 800, 9700AV Groningen, the Netherlands\\
$^{5}$STAR Institute, Quartier Agora - All\'ee du six Ao\^ut, 19c B-4000,  Li\`ege, Belgium\\
$^{6}$Centro de Astrobiolog\'ia (CSIC-INTA), Carretera de Ajalvir km. 4, Torrej\'on de Ardoz, E28850, Madrid, Spain \\
}

\date{Accepted XXX. Received YYY; in original form ZZZ}

\pubyear{2021}

\begin{document}
\label{firstpage}
\pagerange{\pageref{firstpage}--\pageref{lastpage}}
\maketitle

\begin{abstract}
The radio quasar luminosity function exhibits an upturn around $L_{6\rm\:GHz}=10^{23}$ W Hz$^{-1}$  that is well-modelled by a star-forming host galaxy population. This distribution leads some authors to cite star formation as the main radio emission mechanism in so-called radio-quiet quasars (RQQs).  Understanding the origin of RQQ radio emission is crucial for our understanding of quasar feedback mechanisms -- responsible for the regulation of star-formation in the host galaxy --  and for understanding galaxy evolution as a whole. By observing RQQs that have been magnified by strong gravitational lensing, we have direct access to the RQQ population out to cosmic noon, where evidence for twin mini-jets has recently been found in a sub-\textmu Jy RQQ. Here we present radio observations of two lensed RQQs using the VLA at 5~GHz, the latest objects to be observed in a sample of quadruply-imaged RQQs above -30$^{\circ}$. In SDSS~J1004+4112 we find strong evidence for AGN-related radio emission in the variability of the source. In PG~1115+080 we find tentative evidence for AGN-related emission, determined by comparing the radio luminosity with modelled dust components. If confirmed in the case of PG~1115+080, which lies on the radio--FIR correlation, the result would reinforce the need for caution when applying the correlation to rule out jet activity and when assuming no AGN heating of FIR-emitting dust when calculating star formation rates. Our programme so far has shown that two of the faintest radio sources ever imaged show strong evidence for AGN-dominated radio emission.

\end{abstract}




\section{Introduction}

Strong gravitational lens systems involve the multiple imaging of background galaxies or quasars, typically by a foreground galaxy close to the line of sight (see \citealt{2002LNP...608....1C,2006glsw.conf.....M,2010CQGra..27w3001B,2013BASI...41...19J} for reviews). Since the deflection of light is dependent on the distribution of matter, modelling of strong lens systems can map out dark matter in the lensing galaxy, variously probing large  \citep{2002ApJ...575...87T,2003MNRAS.343L..29T,2004ApJ...611..739T,0004-637X-752-2-163,0004-637X-777-2-98,0004-637X-800-2-94} and small \citep{1998MNRAS.295..587M,2001ApJ...563....9M,2004ApJ...610...69K,2012Natur.481..341V,2017MNRAS.471.2224N} scales of dark matter substructure.   Moreover, the consequent magnification of the background object allows us to study it at some combination of higher signal-to-noise and higher resolution (e.g. \citealt{2006A&A...451..865C,spingola2019gravitational}). In particular this allows studies with radio interferometers of objects with intrinsically faint radio emission, whose detailed investigation would otherwise only become possible with next generation telescopes such as the Square Kilometre Array (SKA, \citealt{2015aska.confE..87O}). By using strong lenses to study the faint radio emission from radio-quiet quasars (RQQs), we can obtain direct evidence to address one of the oldest problems in radio astronomy: the radio life of quasars.

Although the first quasars were discovered by identifying strong radio sources associated with highly redshifted optical point-sources \citep{1963Natur.197.1040S}, quasars only rarely appear to exist in a ``radio-loud'' state, with more than 90\% \citep{1965ApJ...141.1560S} displaying only weak radio emission. Understanding the physical mechanism for this weak emission is crucial for our understanding of galaxy evolution as a whole\footnote{The distinction between radio loud quasars RLQs and RQQs is conventionally made using the ratio $R$ of 5~GHz and optical B-band monochromatic luminosities, the cut being made at $R$=10}. Observations of radio-loud quasars (RLQ) reveal kiloparsec-scale jets and lobes resulting from the collimation of galactic gas and dust shredded during its accretion onto a central supermassive black hole (SMBH).  These powerful jets are thought to regulate gas cooling and subsequent star-formation within the host galaxy, constituting a form of ``AGN feedback'' during what historically has been referred to as the ``radio'' or ``maintenance'' feedback mode \citep{2006MNRAS.365...11C,2012MNRAS.421.1569B}. Evidence for corresponding trajectories of star-formation and AGN activity is summarised by  \cite{2014ARA&A..52..415M}. During the so-called ``quasar'' feedback mode, on the other hand, luminous jets are absent; instead, high velocity winds are thought to regulate star formation by driving gas from the galaxy. The radio emission of RQQs, therefore, can trace the processes responsible for this feedback. Determining the mechanism of this radio emission -- whether it is the result of suppression or, alternatively, activation of star formation -- is key to understanding this stage of galaxy evolution.

The question of the origin of radio emission from RQQs has been active for decades: either the emission is the result of a scaled down version of the classic radio jet, is due to ongoing star-formation in the host galaxy\footnote{Indeed, some \citep{2017NatAs...1E.194P} have proposed that the RLQ/RQQ distinction be abandoned altogether and replaced by the designation of ``jetted'' and ``unjetted'' sources }, or is the result of both. Early optically-selected samples of quasars found a bimodality in the distribution of luminosities \citep{1989AJ.....98.1195K}.  A more recent National Radio Astronomy Observatory (NRAO) Jansky Very Large Array (VLA) Sky Survey (NVSS) sample hinted at a large population of \textmu Jy level quasars \citep{2013ApJ...768...37C}. Both results imply a quasar model consisting of two distinct populations, invoking star-formation as the radio emission mechanism of the radio-quiet population. Most recently, \cite{2020MNRAS.492.5297M} used a Bayesian stacking approach to fit quasar radio luminosity functions below the radio detection threshold for a sample of optically-selected quasars, finding a distinct upturn below log$_{10}$[L$_{1.4}$/W Hz$^{-1}$]$\approx$24.8 and coinciding with luminosity ranges where star-forming galaxies are expected to start to dominate. Further, a study of a large number of radio-quiet, gravitationally lensed quasars found that many of them lie on the radio--far-infrared (radio--FIR) correlation occupied by star-forming objects, suggesting that a large component of the radio emission is due to star formation \citep{hannah2,2018MNRAS.476.5075S}. This result is consistent with earlier work by \cite{2013MNRAS.436.3759B,2015MNRAS.453.1079B}, and \cite{doi:10.1093/mnras/251.1.14P}, who first invoked the use of the FIR to make this distinction. More recently, \cite{gurkanrefId0} and \cite{2021MNRAS.506.5888M} have used data from the Low Frequency Array (LOFAR) Two-Metre Sky Survey (LoTSS; \citealt{2019A&A...622A...1S}) to find a lack of bimodality at low radio frequencies but evidence for low-luminosity quasars being dominated by star formation. The authors of both studies suggest that the jet launching mechanism operates in all quasars but with different powering efficiency.

On the other hand, there is some evidence that the same AGN engine is in operation in all quasars, but for RQQs is significantly reduced in power. \cite{1998AAS...19311004B}, for example, found strong evidence for jet-producing central engines in eight objects from a sample observed using the Very Long Baseline Array (VLBA), and \cite{2006A&A...455..161L} found in a sample of 14 low redshift radio–quiet quasars radio structures that can be interpreted as jet–like outflows. \cite{2016A&A...589L...2H} used VLBA observations of intermediate luminosity RQQs to show that the radio emission of at least some RQQs is dominated by AGN activity, and \cite{2016MNRAS.455.4191Z} have found that the star formation rate in a sample of 300 quasars is insufficient to explain the observed radio emission, by an order of magnitude. Others have used source variability \citep{2005ApJ...618..108B} and excess emission from the radio--FIR correlation to rule out star-formation as the dominant emission mechanism \citep{2017MNRAS.468..217W}. Alternative radio emission mechanisms have been suggested, such as magnetically-heated coronae \citep{2018arXiv181010245L,2021arXiv210607783R}, radiatively-driven shock fronts  \citep{2014MNRAS.442..784Z}, and  optically-thin bremsstrahlung emission in the quasar core \citep{2007ApJ...668L.103B}. 

We can use the magnification provided by strong gravitational lenses to make direct study of a sample of RQQs, resolving the radio emission in each case. This method allows us access to the very faintest radio sources at the very highest resolution, with intrinsic flux density values around the 1 \textmu Jy level.  VLA observations of four quadruply-lensed RQQs have detected all four sources, with intrinsic flux densities in the range 1-5 \textmu Jy \citep{2015MNRAS.454..287J}. Follow-up e-Multi-Element Radio Linked Interferometer Network (e-MERLIN) and European VLBI Network (EVN) observations of lensed RQQ HS~0810+2554 \citep{har19} allowed us to resolve radio structures to the sub-pc scale. The images revealed an AGN emission mechanism in the form  of twin jets on opposing sides of the optical core, in what would be, if unlensed, the faintest radio source ever to be observed. Follow-up Atacama Large Millimeter Array (ALMA) and VLA observations of SDSS~J0924+0219, on the other hand, found a radio-emitting region coincident with a molecular disk and of the same apparent size, suggesting star-formation as the main cause of radio emission in this case, which lies on the radio--FIR correlation \citep{Badole_2020}. In the strongly-lensed RQQ RXJ~1131$-$1231, no compact radio sources were detected using EVN observations \citep{2008arXiv0811.3421W} but there is evidence of turbulent gas, again consistent with star-forming activity \citep{paraficz2018alma}.

The picture emerging from this directly-observed sample is mixed, with observations to date consistent either with star-formation or jet activity, or perhaps both. Further, \cite{paraficz2018alma} suggest that star-formation is dependent not on AGN activity but instead on host gas morphology. In order to refine the picture and collect more evidence, we have observed two more quadruply-lensed RQQs using the VLA. We also use our observations to expand the sample of quadruply-lensed radio quasars available for dark-matter sub-structure studies. Perturbations in the distribution of the lensing mass manifest in flux ratios that are inconsistent with predictions from smooth mass models. In the optical, the highly compact size of the accretion disk renders quasars sensitive to so-called microlensing by a compact object, such as a planet or star. Microlensing has only been detected tentatively in the radio \citep{2000A&A...358..793K}, and is considered unlikely due to the greater typical size of a radio source than the ~\textmu as-size microlensing Einstein radius \citep{Dalal_2002}. Radio observations of lensed quasars are therefore thought more reliably to betray the presence of  milliarcsecond- (mas) scale dark matter substructures along the line of sight -- in a phenomenon known as millilensing \citep{2001ApJ...563....9M} -- predicted by the Cold Dark Matter model. While we focus mainly on lensed source astrophysics in this paper, we present preliminary findings on the lensing mass structures, which will be investigated in more detail in a dedicated study. Sections 2 and 3 of this paper present our new observations, their results, and implications for the origin of RQQ radio emission in each source and the mass distribution of the lens. Section 4 presents our conclusions within their wider contexts. Throughout this paper we assume a standard flat cosmology with $\Omega_m=0.31$ and H$_0=67.8$ km s$^{-1}$ Mpc$^{-1}$ (Planck Collaboration XIII \citeyear{2016A&A...594A..13P}).

\section{SDSS~J1004+4112}

SDSS~J1004+4112 \citep{inada2003gravitationally}, is a giant lens system produced by a galaxy cluster lying at redshift $z=0.68$. In the optical, several background objects appear to be distorted by the intervening mass, producing  $\sim$10 arcsecond-scale arcs and filaments of lensed galaxies \citep{sharon2005discovery}, in addition to multiple point-like images of a lensed quasar. The quasar, at a redshift of  $z = 1.734$, is lensed into five images, with four magnified images in a fold configuration separated by a maximum of 14.62 arcseconds, and a fifth, de-magnified image embedded slightly West of the central galaxy in the cluster \citep{2005PASJ...57L...7I,inada2008spectroscopic}. The quasar source displays intrinsic variability in the optical (see e.g. \citealt{2008ApJ...676..761F,fian2016size}), resulting in time delays of years between the more distant components.

 \begin{figure*}
  \begin{tabular}{cc}
      \includegraphics[trim={1.6cm 3cm 0.9cm 6.4cm},clip,width=1\columnwidth]{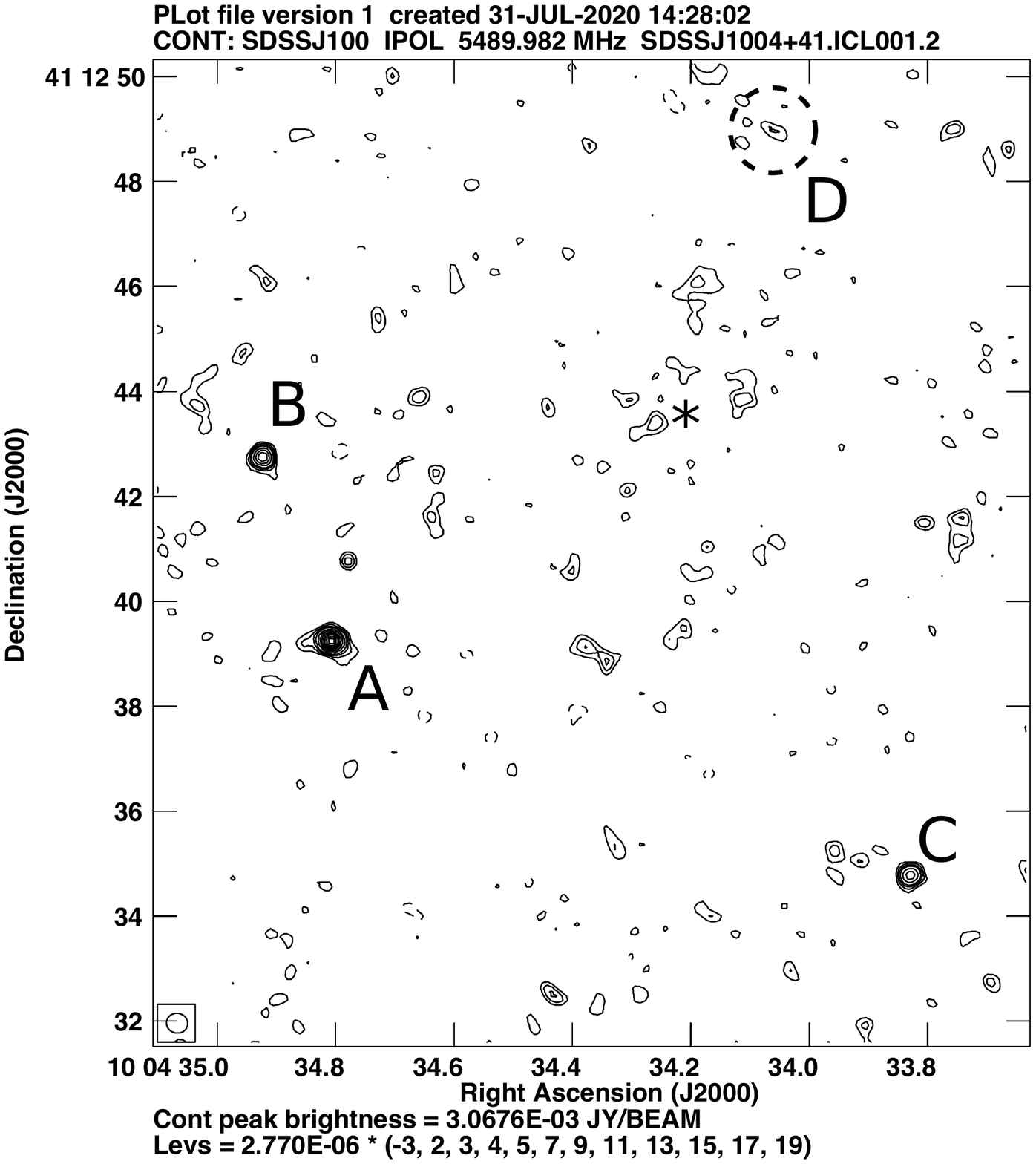} 
      &\includegraphics[trim={1.6cm 3.4cm 1.1cm 5.8cm},clip,width=1\columnwidth]{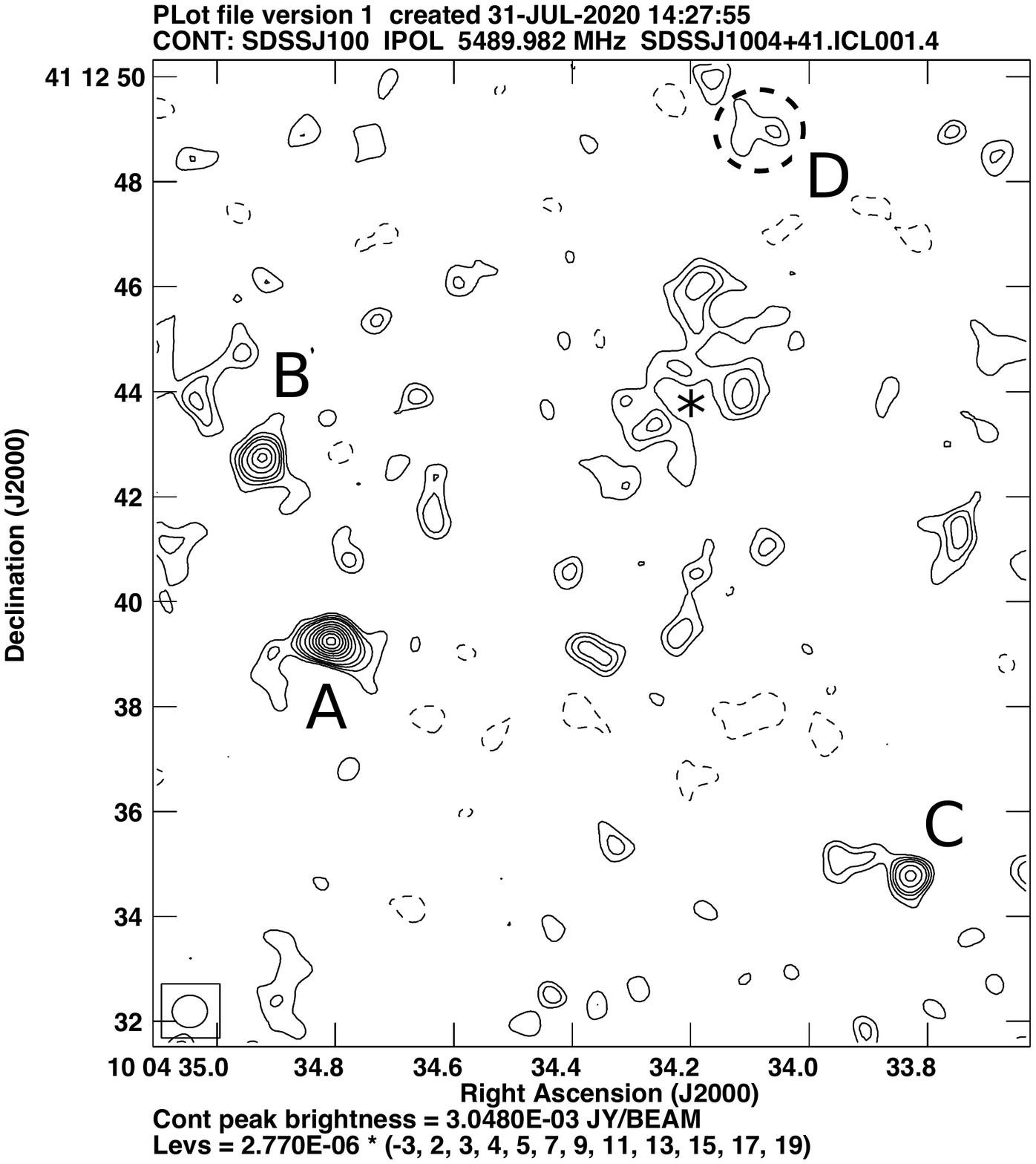} \\
      \end{tabular}
  \caption{Left: The final 5~GHz A-configuration VLA map of J1004+4112, produced using natural weighting and with a restoring beam at full width of half maximum (FWHM) $405\times377$ mas at a position angle of $77^{\circ}$. The asterisk indicates the approximate location of the lensing galaxy in the optical and the dashed circle is centred on the expected location of component D. Right: Map of the same field as Fig.~\ref{1004crop} but produced using a $u-v$ tapering scheme. A Gaussian function was used to weight down longer baseline lengths, with a distance to 30\% of the Gaussian height of 300 k$\lambda$. The restoring beam used a FWHM of $670\times 620$ mas at a position angle of $-85^{\circ}$.}

 \label{1004crop}
\end{figure*}

The first radio observation of SDSS J1004+4112  \citep{2011ApJ...739L..28J} used the improved sensitivity of the newly expanded VLA at 5~GHz in C configuration to produce an image of the system at a resolution of 3.5 arcsec. The resulting radio map shows a clear detection of four magnified quasar image components, with the brightest components, A and B, appearing to merge. Component E was not detected. The system has recently been observed at 144 MHz, within LoTSS data release 2. Resulting images made detections of components A and B with flux densities of 370 \textmu Jy and 272 \textmu Jy, respectively, while C and D were not detected \citep{2021MNRAS.tmpL..33M}.  Using the relation from \cite{2017MNRAS.469.3468C} and under an assumption that all of the emission originates from star-forming activity, the host galaxy star formation rate (SFR) would be  $5.5^{+1.8}_{-1.4}$~M$_\odot$ yr$^{-1}$. In order to resolve individual 5 GHz  components to greater detail and to determine the radio emission origin, further VLA C~band observations have been made, using longer baselines and thus higher resolution. 

\subsection{Observations and results}

SDSS~J1004+4112 was observed on 2016 November 9 and 21 using the VLA at 4.5--6.5~GHz (C~band) in the A configuration. The total on-source integration time was 2.5 hours.  A nearby point source, J0948+4039, was used for phase referencing. Data reduction was performed using calibration and imaging routines from the Astronomical Image Processing  System  ({\sc aips})  software  package  distributed  by  NRAO\footnote{\tt http://www.aips.nrao.edu}.  While 3C286 was observed in order to set the overall flux, its observation was unintentionally divided into two spectral windows rather than the 16 used for the target and phase calibrator. The overall flux was therefore set by using the phase calibrator, assuming a flat spectral index and 1.2~Jy -- according to values from the National Aeronautics and Space Administration (NASA) extragalactic database\footnote{\tt https://ned.ipac.caltech.edu} -- and inserting flux values for each spectral window using {\sc aips} task {\sc setjy}. After applying delay and phase solutions to all sources and using {\sc aips} task {\sc getjy} to bootstrap the flux scale, the calibrated 3C286 observation was mapped and its flux compared with available models to confirm that the correct flux scale had been set \citep{2017ApJS..230....7P}, agreeing within the measurement uncertainties.

After applying all solutions to the target, the two epochs of observations of SDSS~J1004+4112 were combined in the $u-v$ plane and imaged. A relatively bright source close to the target permitted self-calibration to be performed, refining the phase and amplitude solutions and improving the noise levels in the final map. The final map was produced using natural weighting, with a restoring beam FWHM of $405\times374$ mas at a position angle of $79.72^{\circ}$.

\begin{figure}

      \includegraphics[trim={1.6cm 5cm 1.5cm 5.9cm},clip,width=1\columnwidth]{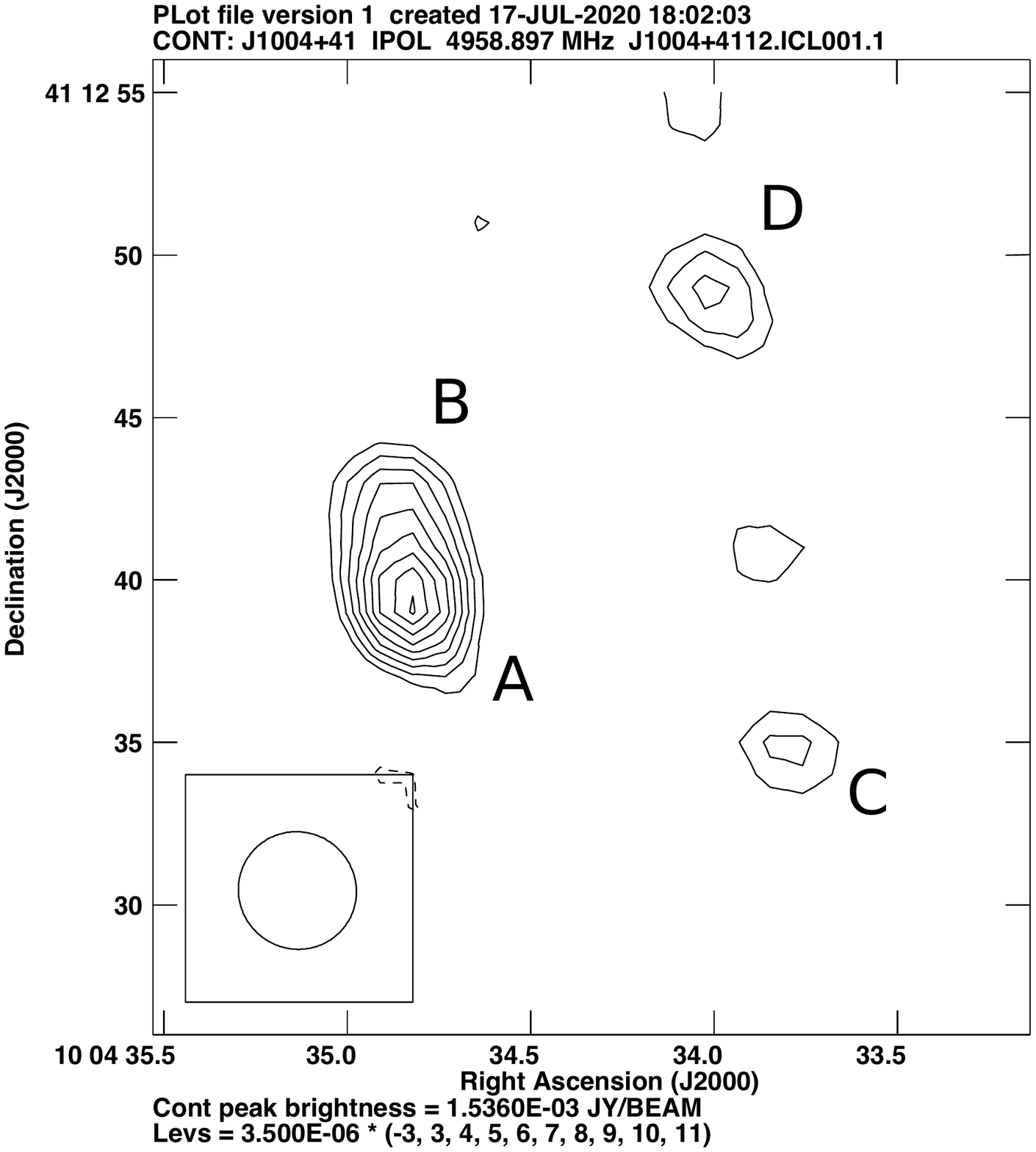} 

  \caption{The previous observation of SDSS~J1004+4112, taken six years earlier in the VLA C-configuration. The map was produced using robust weighting and with a circular beam FWHM of 3.6 arcsec. All four lensed images are visible, with the brightest two exhibiting a merging morphology at this resolution.}

 \label{10042010}
\end{figure}

The previous observation, taken on 2010 October 15 in the VLA C-configuration, was also re-analysed for uniformity and to check the flux scale. This was set using 3C286, and the flux scale was checked by producing calibrated images of 3C286 using the same calibration applied to SDSS~J1004+4112. The final map, produced with robust weighting, has a beam of 3.6 arcsec (Fig.\ref{10042010}).

\begin{table}
	\centering
		{\footnotesize
	\begin{tabular}{lcccc} 
	
		   	\hline
        	\hline\noalign{\smallskip}
    	Cpt. &	$405\times377$ mas &  $670\times 620$ mas  & 3.95$\times$3.69\arcsec & $\mu$ 	\\
    	 &	Nov 2016 & Nov 2016 & Nov 2010 &   \\ \hline\noalign{\smallskip}
			   
			   A&$56.1\pm2.8$&$54.1\pm3.6$&$64\pm8$&29.7\\
			   B&$35.0\pm2.8$&$34.6\pm3.6$&$39\pm8$&19.6\\
			   C&$31.7\pm2.8$&$31.6\pm3.6$&$30\pm8$&11.6\\
			   D&$10.1\pm2.8$&$10.1\pm3.6$&$33\pm8$&5.8\\
			   			   
		\noalign{\smallskip}\hline
		  	\hline \noalign{\smallskip}
	\end{tabular}
        }
    	\caption{Flux density values in \textmu Jy for each lensed component in the VLA 5~GHz observations of J1004+4112, measured using {\sc aips} task {\sc jmfit}. Two sets of values are measured for the observations made using the VLA A configuration: one from the image produced using natural weighting (column 2), and one from the image produced after using a Gaussian function to weight down longer baseline lengths (see Section~\ref{1004radioorigin}), with a distance to 30\% of the Gaussian height of 300 k$\lambda$ (column 3). The measurements made by \citet{2011ApJ...739L..28J} from C band observations using the VLA at C configuration are also presented, for comparison (column 4). Magnification $\mu$ values predicted by the model of \citet{oguri2010mass} are included for reference.}
        \label{1004fluxes}
\end{table}

 \begin{figure}

      \includegraphics[trim={0.5cm 0.3cm 1.5cm 1.3cm},clip,width=1\columnwidth]{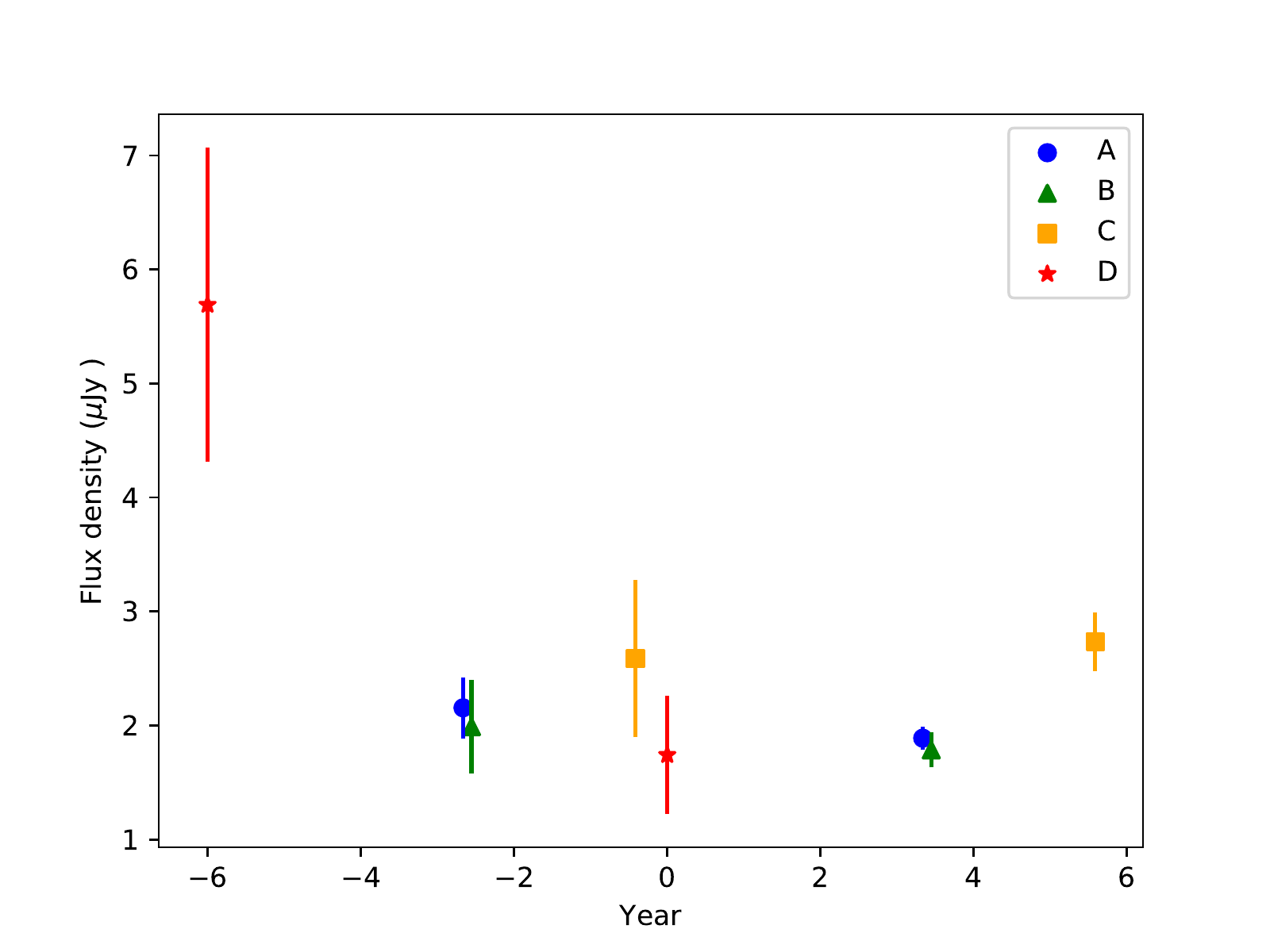} 

  \caption{Flux density values of each lensed image of SDSS J1004+4112 plotted over time, using data from the 2010 and 2016 VLA 5 GHz observations.  The arrival times observed by  \citet{Fohlmeister_2007} between A, B and C and predicted by \citet{oguri2010mass} for D have been used to indicate the time of emission of each component in both epochs. The flux density of each component has been divided by the magnification factors of \citet{oguri2010mass}.}

 \label{times}
\end{figure}

The final A-configuration VLA map of SDSS~J1004+4112 is presented in Fig.~\ref{1004crop}. The off-source root mean squared (r.m.s.) noise is measured at 2.8~\textmu Jy/beam. The flux density values associated with each component were obtained by using {\sc aips} task {\sc jmfit} to fit single Gaussian components in the image plane. The values and their uncertainties are reported in Table~\ref{1004fluxes}. Components A, B and C are clearly visible above the noise, with flux values and ratios which agree with the 2010 C~configuration image, within the errors. A tentative detection at image D is made. No detection is made in the optical location of image E. The flux density of image D is inconsistent with the 2010 measurement, which is significantly higher (see Figure~\ref{10042010}).  All components are unresolved.

\subsection{Structure of the lensing cluster}
\label{1004structure}

SDSS J1004+4112 has undergone mass modelling of varying complexity in the literature \citep{inada2003gravitationally,2004ApJ...605...78O,williams2004models,liesenborgs2009non}.  Most recently,  \citet{oguri2010mass} has modelled the system as a parameterised composite mass distribution in order to account explicitly for individual galaxies in addition to a smooth dark matter halo and an external shear component. The authors used constraints from measured time delays of lensed quasar components, as well as their positions and flux densities and those of additional lensed galaxies. The model is found to be in agreement, within photometric fitting errors, with the lensed quasar optical flux ratio constraints measured in April 2004 by \cite{2005PASJ...57L...7I}.  Ratios of the magnification values predicted by the model at each lensed component are in agreement with the flux ratios of components A, B and D in the VLA 5~GHz 2016 observation. Magnification ratios involving component C, however, are inconsistent with those in the radio map, with component C appearing to be brighter by at least one third than the mass model would predict. The flux ratios of D in the earlier epoch are also inconsistent with the modelled magnification values, being anomalously bright by a factor of at least 2:  intriguingly, component D appears to have weakened since the 2010 observation, while components A, B and C show no variation.

Several mechanisms could explain the anomalous brightness values observed at components C and D. Millilensing, due to sub-halo sized dark matter structures in the lensing cluster or along the line of sight, cannot be ruled out. While optical observations of this system can be well-modelled without invoking substructure components, it is possible that the radio emission originates from a source that is offset from the optical quasar core and therefore that the lensed emission has traversed different regions of the intervening mass distribution. This scenario is consistent with flux measurements taken at component C from both epochs, although the relative uncertainties of the 2010 values are too large to make definitive statements, being consistent also with the magnification ratios from the optical model. The timescale of variability due to lensing -- being proportional to the Einstein radius of the lens -- increases with the square root of the lensing mass. This precludes detection of any millilensing-induced variability over our observational timescale and rules out millilensing of a static source as the cause of the variability of component D. It is possible, however, that the source is not static and is instead moving with respect to the lensing caustics of a millilensing object, which could result in the flux density variation observed at image D (see Section~\ref{1004radioorigin}). 

Another explanation is that components C and D are undergoing microlensing events, which have persisted during both the 2010 and 2016 observations. While considered unlikely in the radio due to source sizes typically extending over regions larger than microlensing caustics, VLBI observations of radio-loud AGN have found jet structures of sizes down to the \textmu as scale and below (see e.g. \citealt{lister2016mojave} and \cite{jorstad2016vlba} for recent studies). Furthermore, imaging of twin mini jets in quadruply-lensed RQQ HS~0810+2554 finds jet components no larger than 0.27 pc at z = 1.51. These spatial scales suggest that the variability of components C and D is possibly due to the microlensing of a small jet structure such as that typically found within $\sim$mas of the central SMBH \citep{Blandford_2019}. The locations of C and D, far from the central galaxy in the lensing field, occupy apparently sparse environments.  Tidal interactions within the cluster, however, would likely increase the abundance of compact objects between galaxies. Furthermore, \cite{2020A&A...643A..10H} have shown that, if the assumption of smoothly distributed dark matter is relaxed, microlensing events in J1004+4112 could be explained by assuming that the cluster’s dark matter is in the form of primordial black holes.

Finally, it is possible to avoid the need to include additional lensing structures if instead intrinsic source variability is considered.  \cite{Fohlmeister_2007} have performed optical monitoring over the four year period 2003-2007 to demonstrate intrinsic variation of the source of at least 0.5 mag over the time scale. The variation allowed the authors to measure time delays between lensed components of the system, which arrive in the order C-B-A-D, with C-A and B-A delays of 821.6$\pm$2.1 days and 40.6$\pm$1.8 days, respectively. The mass model of \citeauthor{oguri2010mass} has predicted a D-A delay of 1218 days, while \citeauthor{Fohlmeister_2007} obtain a lower limit of 1250 days and median 2000 days. Using the respective time delays to plot 5~GHz image flux densities against time of emission from the source (Fig~\ref{times}), the anomalous flux density values of components C and D could be explained by intrinsic variation of the source; for their flux ratios with other components to be consistent with the model of \citeauthor{oguri2010mass}, the source would be required to have weakened by at least a half and later brightened by at least a third over three and two year time scales, respectively. This is possible for a source size smaller than 0.6 pc.

There is no convincing detection of a de-magnified fifth lensed image (component E) in the 5 GHz radio maps. Such a detection would have been unlikely given the low flux densities of the magnified images, despite the relatively shallow mass profiles of cluster lenses resulting in a more modest de-magnification of the central image  \citep{wallington1993influence,rusin2001constraints,2004Natur.427..613W,2016MNRAS.459.2394Q}. Using the predicted magnification value of $\mu_E = 0.16$ from the model of \citeauthor{oguri2010mass} in combination with the flux density and magnification value of component A, a flux density of 0.3 \textmu Jy would be expected at the location of the fifth image: significantly below the 2.8 \textmu Jy/beam r.m.s. noise.

Given that the observational results at 5~GHz are consistent with scenarios that do and do not feature  micro- and millilensing, events, it is not possible to make a conclusive statement on the presence of small scale lensing structures in the system. It is possible, however, to use the data to draw tentative conclusions on the nature of the source radio emission.

\subsection{Origin of the RQQ radio emission}
\label{1004radioorigin}

The flux densities measured at component C are consistent with all scenarios presented in section~\ref{1004structure}. The variation and anomalous flux density of component D, however, is inconsistent with a scenario involving the millilensing of a static source, which would not be observable over the 6 year timescale between epochs. Ruling this out, therefore, leaves the possibilities  of microlensing, of millilensing of a moving source, and of intrinsic variability of the source, all of which have the requirement that the source is either very compact or is moving. The scenarios of microlensing and intrinsic source variability both require an emitting region smaller than $\sim1$pc in size, within which either a radio jet or a supernova event (see e.g. \citealt{10.1111/j.1365-2966.2004.08004.x}) could reside. Using the $\alpha^{\rm 144 MHz}_{\rm 5GHz}$ spectral index of -0.46$\pm0.05$ measured by \cite{2021MNRAS.tmpL..33M} and an intrinsic flux density of $\sim$2 \textmu Jy estimated using the lens model of \citeauthor{oguri2010mass}, the rest-frame luminosity of SDSS J1004+4112 can be estimated at $L_{\rm 5 GHz}\sim2.4\times 10^{22}$ W\,Hz$^{-1}$. Given that the maximum supernova luminosity at 5~GHz is of order $\sim3\times10^{20}$ W\,Hz$^{-1}$ \citep{1987AJ.....94...61R}, supernova activity is very unlikely to be responsible for the radio emission in this case.

For the millilensing scenario, the requirement of a moving source could be fulfilled in the case of an AGN-launched jet, especially in the case of a jet undergoing apparent superluminal motion. For example, according to geometry, a jet travelling at a velocity of 0.95c at an angle 20 degrees from the line of sight would project an apparent transverse motion of $\sim$3c , which could traverse angular distances of over 0.5 mas in the space of 6 years: far enough to move behind a millilensing object in the lensing cluster. SDSS~J1004+4112, a broad line quasar, is likely to be inclined at a shallow angle from our line of sight. While any jets present in this faint source must be very low powered, relativistic velocities may still be present.

Two further possible causes of the lower flux of component D in the 2016 image could be the result of scattering by intervening plasma, which would result in a broadening of the image \citep{2003MNRAS.338..599B}, or of the presence of another radio source in the vicinity of image D,  such that the higher resolution observation would resolve out or de-blend, respectively, some of the source flux. In order to investigate the possibility of scattering, the data were re-imaged using heavy tapering on the long baselines. A map produced using a Gaussian function to weight down longer baseline lengths, with a distance to 30\% of the Gaussian height of 300 k$\lambda$, is presented in Fig.~\ref{1004crop}.  This map shows no increase in the flux density of component D with respect to the full baseline image. Inspection of Hubble Space Telescope (HST) images from \cite{oguri2010mass} finds no significant nearby optical source.  Scattering and deblending from a nearby source are therefore ruled out, leaving the possibilities either of variability of the radio source or of milli- or microlensing along the line of sight as the cause of the lower flux.  

While various explanations are possible for the disagreements of flux values at D, all possible scenarios require that the radio emission originates from a highly compact and/or moving source. Based on these measurements, therefore, we suggest that the origin of radio activity in this source is an AGN core or jet.  The spectral index $\alpha^{\rm 144 MHz}_{\rm 5 GHz}$ of -0.46$\pm0.05$ obtained by \cite{2021MNRAS.tmpL..33M} is flatter than is typical either for AGN jet or for star-forming regions, but is steeper than typically found for radio AGN core emission \citep{2014AJ....147..143H}, where significant synchrotron self-absorption is expected. It is possible that the emission is the combined result of core and jet activity. On the other hand, recent studies using LoTTS have found that the spectral index of lower brightness AGN jets appears to tend towards flatter values \citep{lotssagn}. 

Component D has also been observed to vary in a consistent way over the same time period in the in C IV broad line region emission \cite{Popovi__2020}. Additionally, while the 2010 VLA epoch was observed a month after last epoch of \cite{fian2016size}, component D appears to be on a downward trend in the optical at that point in time. A scenario involving related AGN processes driving a mutual intrinsic variation in the radio, optical and C IV line emission is therefore an attractive one. No IR photometric measurements are recorded in the literature, with a lack of detection in the AKARI all-sky catalogue \citep{2007PASJ...59S.369M} providing an upper limit with respect to the radio--FIR correlation \citep{2010A&A...518L..31I}, quantified by parameter $q_{\rm IR}$, of 3.9.

A lower limit to the source brightness temperature $T_{\nu}$ can be calculated as follows: 

\begin{equation}
T_{\nu} = 1222\times \frac{I}{\nu^2\theta_{\rm maj}\,\theta_{\rm min}},
\end{equation}

where $\nu$ is the observing frequency in GHz, $I$ is the source brightness intensity measured in mJy beam$^{-1}$, and  $\theta_{\rm maj}$ and $\theta_{\rm min}$ are the major and minor axis FWHM beamwidths in arcseconds. Projecting component A back to the source plane using the magnification factor from \cite{oguri2010mass}, and using the flux density after applying the standard radio K-correction, $(1+z)^{\alpha-1}$, with the spectral index $\alpha^{\rm 144 MHz}_{\rm 5 GHz}$ of -0.46$\pm0.05$ obtained by \cite{2021MNRAS.tmpL..33M},   we arrive at a lower limit of $T_{5.5\rm GHz} = 240$~K. The resolution afforded by Very Long Baseline Interferometry (VLBI) would provide much greater constraints to the brightness temperature,  with a detection alone confirming AGN activity as the radio emission mechanism. VLBI maps may also distinguish between a core-jet and jet-only structure.

\section{PG~1115+080}

Discovered by \citet{1980Natur.285..641W}, PG~1115+080 is a lens system in a merging fold configuration, situated within a galaxy group. The lensed source is a broad absorption line (BAL) quasar with a redshift of $z = 1.722$, and the lensing galaxy lies at $z = 0.311$ \citep{1997AJ....114..507K} and has an Einstein radius of about 1 arcsec. A faint arc of unobscured star-formation in the host galaxy is visible in the near-infrared and optical \citep{1998ApJ...509..551I,sluse1115,10.1093/mnras/stz2547}. The merging pair of quasar images shows a flux ratio anomaly in the optical, which has varied over time (e.g. \citealt{1997ApJ...475L..85S,cour,1998ApJ...509..551I,0004-637X-648-1-67}). \citet{2010MNRAS.406.2764T} collated past values of the ratio and performed additional new observations, finding that the A/B ratio showed a distinct trend over a timescale of 26 years since the object's discovery, beginning at approximately unity and falling to 0.5 before rising to approach unity again. The authors concluded that the steady fall and rise was indicative of microlensing in the lens galaxy.  Observations by \citet{0004-637X-648-1-67} found a stronger flux anomaly still in the X-ray, with a value of 0.2. That the X-ray anomaly is more extreme than in the optical further supports a microlensing scenario, since a stronger effect would be expected from the smaller X-ray source size.  The authors argue that the higher ratio in the optical implies an optical region $\sim$10-100 times larger than expected from a thin accretion disk mode.

\subsection{Observations and results}

PG~1115+080 was observed on 2016 November 23rd using the VLA at 4.5--6.5~GHz (C~band).  The total on-source integration time was approximately 40 minutes. Calibration and imaging was performed manually using {\sc aips},  with 3C286 observed in order to set the overall flux, and nearby point source J1116+0829 used as the phase reference source. The final map was produced using natural weighting, with a restoring beam FWHM of $443\times370$ mas at a position angle of $-15.4^{\circ}$.

\begin{figure}

      \includegraphics[trim={1.6cm 6cm 1.cm 6.cm},clip,width=1\columnwidth]{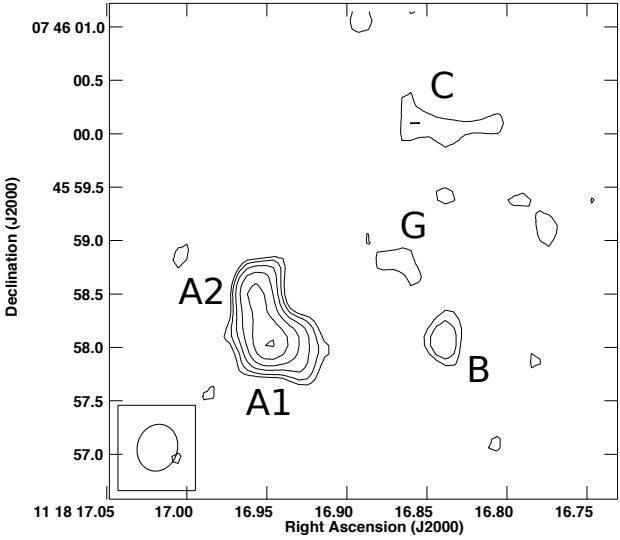} 

  \caption{The final 5~GHz VLA map of PG~1115+080, imaged using a $443\times370$ mas restoring beam at a position angle of $-15.4^{\circ}$. Contours begin at 2$\;\sigma$ at 10 \textmu Jy/beam and increase by factors of $\sqrt{2}$. A clear detection is made, and a flux anomaly is noted between lensed image components A1 and A2. The lensing galaxy is also seen.}

 \label{1115ann}
\end{figure}

The final radio map of PG~1115+080 is presented in Fig.~\ref{1115ann}. All four lensed components are visible, and the lens galaxy is tentatively detected in the centre. The off-source r.m.s. noise is 5.1~\textmu Jy/beam. The lensed components hint at some extension to the lensed source: this is noted in the shape of merging components A1 and A2 and in the shape of component C.  Table~\ref{1115fluxes} reports the total flux density measurements using {\sc jmfit} to fit Gaussian components to the lensed components in the image plane.

\begin{table}
	\centering
		{\footnotesize
	\begin{tabular}{ccccc} 
	   	\hline
        	\hline\noalign{\smallskip}
    	Comp. &	Flux density (\textmu Jy)& Maj. (mas)&Min. (mas)&PA (deg) \\ \hline\noalign{\smallskip}
			   
	    A1&$ 82 \pm12$&$667\pm 65$&$489\pm48$&$78\pm13$\\
	    A2&$ 42 \pm5$*&$474\pm57$&$343\pm41$&$176\pm15$\\
	    B&$ 20 \pm5$*&$485\pm120$&$344\pm86$&$4\pm28$\\
	    C&$ 35 \pm17$&$853\pm265$&$487\pm151$&$84\pm21$\\	    
	    G&$ 22 \pm13$&$875\pm297$&$359\pm122$&$50\pm14$\\
		\noalign{\smallskip}\hline
        	\hline \noalign{\smallskip}
	\end{tabular}
        }
    	\caption{Flux density values, values of the major and minor axes, and position angles for each component in the VLA C~band observations of PG~1115+080. The measurements and uncertainties were obtained by using {\sc aips} task {\sc jmfit} to fit Gaussian components to the image components. * denotes that the component is unresolved. Position angles are measured East of North. }
        \label{1115fluxes}
\end{table}

\subsection{Unlensed source modelling}

In order to obtain a reconstruction of the background source we used the {\sc visilens} package \citep{hezaveh13a,spilker16a} to fit a model directly to the visibility data. Fitting in the visibility plane has the advantage that all calibrated data are used and avoids the danger of fitting to deconvolution artefacts in the image plane. Using parameterised models of the source and lens, {\sc visilens} performs inverse ray-tracing to populate an image plane with surface brightness values from the source plane, before transforming into Fourier components and interpolating onto the observational $u$-$v$ coordinates. An objective function is constructed using the values of observed, ${V}^{\rm o}_i$ and model, ${V}^{\rm m}_i$ complex visibilities, such that the $\chi^2$ value is equal to 

\begin{equation}
    \chi^2 = \sum_{j}^N\frac{({\rm Re}(V_i^{\rm o})-{\rm Re}(V_i^{\rm m}))^2+({\rm Im}(V_i^{\rm o})-{\rm Im}(V_i^{\rm m}))^2}{\sigma_{i}^2}, 
\end{equation}

\noindent where $\sigma_{i}$ is the error on each visibility, and is calculated by rescaling the relative visibility weights determined by {\sc aips} by the mean difference between successive visibilities on each baseline. {\sc visilens} uses the {\sc emcee} package \citep{2013PASP..125..306F} to perform Markov Chain Monte Carlo (MCMC) analysis, which explores the parameter space to build posterior distributions of model parameters while maximising the log likelihood, ${\rm ln}\ell$. Making the assumption that the errors are Gaussian and uncorrelated, the log likelihood is trivially related to the $\chi^2$ value: ${\rm ln}\ell = -\chi^2/2$.

We modelled the system using a singular isothermal ellipse (SIE) plus external shear to represent the lensing mass, and a single Gaussian profile to represent the source. We fixed the ellipticity $e$ and position angle $\theta$ of the lens, and the magnitude $\Gamma$ and position angle $\Gamma_{\theta}$ of external shear to the values of the model predicted by \cite{2002ApJ...565...17C}. We allowed the lens position $x_{\rm L}$,  $y_{\rm L}$ and mass  $M_{\rm L}$, and source position $x_{\rm S}$, $y_{\rm S}$, flux density  $F_{\rm S}$, major axis  $a_{\rm S}$, axis ratio $b_{\rm S}/a_{\rm S}$ and position angle $\phi_{\rm S}$ to vary. The light profile is parameterised such that half-light radius $R_{\rm eff} = a_{\rm S} \times (b_{\rm S}/a_{\rm S})^{0.5}$. Uniform priors were placed on all parameters. A milliJy source located approximately 2 arcminutes from the phase centre was subtracted from the observed visibility data, before averaging heavily in time and frequency in order to reduce modelling computation time. Experiments to subtract the faint emission of the lensing galaxy from the visibility data were also made, but this did not appear to improve results, possibly because the apparent emission may include deconvolution artefacts. Setting sensible priors for the lensing galaxy position and its mass avoided the inclusion of the lensing galaxy emission in the source model.

\begin{figure*}

      \includegraphics[trim={0cm 0cm 0cm 0cm},clip,width=1\linewidth]{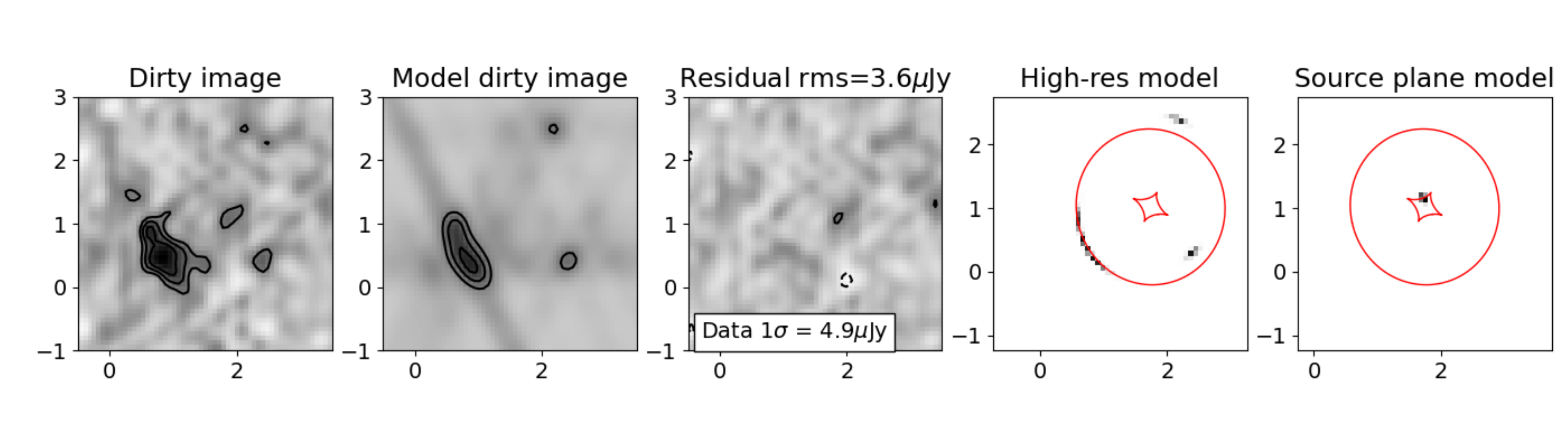} 

  \caption{Results from modelling the 5~GHz VLA observations of PG~1115+080 using {\sc visilens}. From left: the map of observed data convolved with the beam; the map of the modelled data convolved with the beam; residual map; modelled image plane and lensing caustic curves; modelled source plane and lensing caustic curves. Contours are drawn at (3, 6, 9) times the observed image r.m.s. value of 4.9 \textmu Jy/beam for the observed and model maps, and at (-2, 2) times the observed image r.m.s. for the residual map. }

 \label{1115mod}
\end{figure*}

\begin{figure*}

      \includegraphics[trim={6.5cm 5cm 7.5cm 9cm},clip,width=1\linewidth]{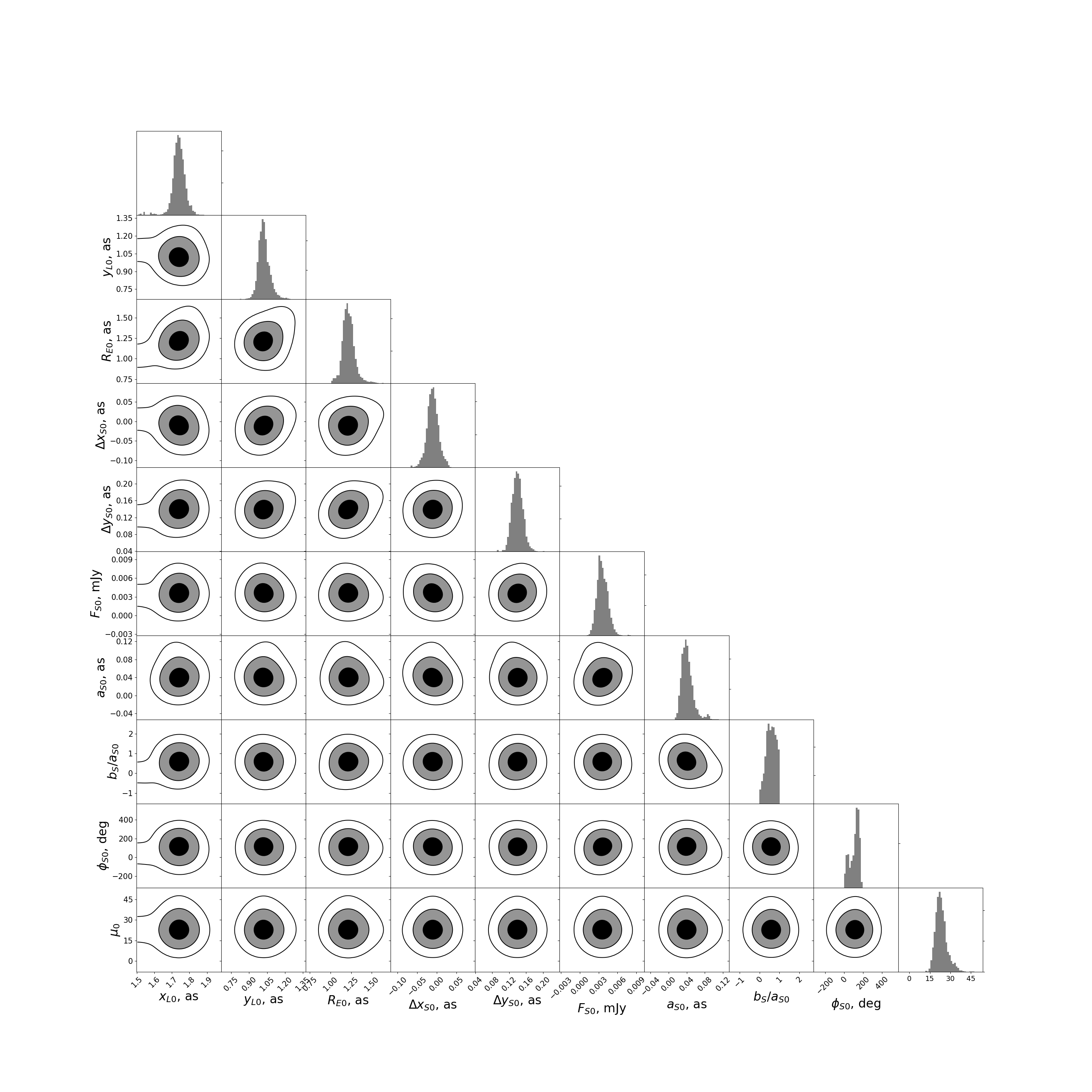} 

  \caption{Marginalised posterior probability distribution functions (PDFs) obtained via {\sc mcmc} analysis of the PG~1115+080 model parameter space within {\sc visilens}. Contour levels drawn at 1, 2 and 3$\sigma$.}

 \label{1115triangle}
\end{figure*}

An image of the best-fit model is shown in Fig.~\ref{1115mod} and  parameter values with associated uncertainties are reported in Table~\ref{1115model}. Fig~\ref{1115triangle} shows the marginalised posterior probability distribution function (PDF) of the modelled parameters. A $\chi_{\rm red}^2=1.1$ represents a reasonable fit to the data. The model predicts an extended source of size 40 mas and 23 mas along the major and minor axes, respectively, representing a physical major axis extent of 348~pc with an effective radius $R_{\rm eff}=120$ pc. An intrinsic source flux density of 4 \textmu Jy is predicted.

   \begin{table}
	\centering
	\begin{tabular}{lcc} 
    	\hline
        	\hline\noalign{\smallskip}
                  
	    Parameter & Value \\
		\hline \noalign{\smallskip}

	    $x_{\rm L}$ (arcsec) & $1.739^{+0.029}_{-0.032}$\smallskip\\
	    $y_{\rm L}$ (arcsec) &$1.020^{+0.039}_{-0.050}$\smallskip \\
        $R_{\rm E}$ (arcsec) & $1.213^{+0.059}_{-0.069}$\smallskip\\
        $x_{\rm S}$ (arcsec) &$-0.010^{+0.013}_{-0.014}$\smallskip\\
		$y_{\rm S}$ (arcsec) &$0.139^{+0.012}_{-0.013}$\smallskip\\
	    $F_{\rm S}$ (mJy) &$0.004^{+0.001}_{-0.001}$\smallskip\\
		$a_{\rm S}$ (arcsec) &$0.040^{+0.010}_{-0.013}$\smallskip\\
		$b_{\rm S}$/$a_{\rm S}$ &$0.586^{+0.026}_{-0.027}$\smallskip\\
		$\phi_{\rm S}$ (deg) &$33.306^{+79.3}_{-32.4}$\smallskip\\
		$\mu$ & $22.760^{+3.393}_{-4.326}$\smallskip\\
        
		\noalign{\smallskip}\hline
        	\hline \noalign{\smallskip}
	\end{tabular}
    	\caption{50th, 16th and 84th percentile model parameters obtained by using the calibrated VLA 5~GHz visibility data to fit a model of the PG~1115+080 system in the $u$-$v$ plane.  The fitting was performed using {\sc visilens } \citep{hezaveh13a,spilker16a}. The lens position $x_{\rm L}$, $y_{\rm L}$ is quoted with respect to the phase centre of 11h18m17.00s +7$^{\circ}$45'57.70" (epoch J2000) and the source position $x_{\rm S}$, $y_{\rm S}$ with respect to the lens. $R_{\rm E}$ is the Einstein radius of the lens, $F_{\rm S}$ is the flux density of the source, $a_{\rm S}$ is the major axis FWHM, $b_{\rm S}$/$a_{\rm S}$ is source axis ratio, $\phi_{\rm S}$ is the source position angle and $\mu$ is the total magnification. Position angles are quoted East of North.}
        \label{1115model}
\end{table}

\subsection{Substructure in the lensing galaxy }

The C~band VLA observations of PG~1115+080 show an A/B peak flux density ratio of $0.51 \pm 0.10$ while -- in the case of an unresolved source -- a ratio of approximately 1 would be expected from the merging fold configuration of the lensed components.   The model prediction of an extended source would appear to explain the flux ratio. The comparison of observed and modelled images in Fig.~\ref{1115mod} illustrates that differential lensing of the source due to its location across different parts of the caustic can be responsible for the seemingly anomalous flux distribution of these observations. This result is consistent with those from the optical, which indicate that the optical flux anomaly in this system is due to microlensing, and also with the mid-infrared \citep{Chiba_2005}, where a flux ratio of 0.93 is found by the authors to be virtually consistent with smooth lens models. Milliarcsecond-scale substructure is not required to explain the observations made to date.

The modelled source size of 40~mas provides a limit on the possible size of substructures in the region of the A/B merger. Using the simple requirement that the Einstein radius of a lensing object must be larger than a background source in order to magnify it and that the object is located at the  same redshift as the macro lens, we determine an upper limit of $M_{\rm ap} =  2.7\times 10^8{\rm M}_{\odot}$ for any Singular Isothermal Sphere (SIS) substructures in the vicinity of the A/B pair, where $M_{\rm ap}$ is the mass enclosed within the aperture defined by the Einstein radius. The simulation results of \cite{10.1046/j.1365-8711.2002.05252.x} showed that an emitting source must be at least an order of magnitude greater in size than the Einstein radius of a lensing object for the object to produce no additional magnification. This places a tighter limit of 4~mas on possible substructure radii, equivalent to a mass $M_{\rm ap} =  2.7\times 10^6\;{\rm M}_{\odot}$.

\subsection{Origin of radio emission in the lensed RQQs}

The unlensed RQQ source is predicted by our model to be relatively extended and quite linear, with a major axis size of $348\pm{103}$ pc. While the model does not put very tight constraints on the axis ratio, the extension seen to the South of merging components A1 and A2 further suggests a linear source morphology. The size and shape of the object is consistent with those of a small and compact radio jet, and is small than often is observed in star-forming regions. For example,   \cite{Badole_2020} performed unlensed source modelling of CO line and radio emission in RQQ SDSS~J0924+0219 to predict a star-forming region 1--2.5~kpc in extent.  On the other hand, the source is comparable in size with some starburst regions, which can range in size from just 0.1 to over 2~kpc  \citep{HinojosarefId0}. Based on source morphology alone, therefore, it is not possible to distinguish between AGN or starburst origin of radio emission. 

We assume a power law dependence of flux density $S$ with frequency $\nu$, $S_{\nu}\propto \nu^{\alpha}$. Using the total lensed 8.4 GHz flux density $S_{8.4 \;\rm GHz} = 153\pm17$ \textmu Jy from \cite{2015MNRAS.454..287J} and the lensed flux density of the four lensed components plus the lensing galaxy emission at 5~GHz,  we find a spectral index of $\alpha = -0.65\pm0.38$. This value makes the assumption that the lensing galaxy emission has the same spectral index dependence and contributes negligibly to the total flux density. The value is again consistent with both AGN and star-forming radio emission (see e.g. \citealt{10.1093/mnras/stx1040} for recent spectral index studies). The beamsize of the A configuration VLA observation only allows us to measure a rest frame brightness temperature lower limit of $T_{5.5\rm GHz} = 7$K, using the flux density value of component A. In order to make further statements on the possible origin of the radio emission, we compare the source brightness and morphology with multi-wavelength values from the literature.

\cite{Chiba_2005} have used the cooled mid-infrared camera and spectrometer (COMICS) attached to the Subaru Telescope to perform mid-infrared (mid-IR) imaging at 11.7~\textmu m of PG~1115+080. The resulting image shows four point-like components of emission, the photometric measurements of which the authors use to identify a clear infrared bump at the rest wavelength of 4.3 \textmu m in the spectral energy distribution (SED) of the source. Limits to the source size derived from lensing magnification factors in the tangential direction suggest that the emission originates from a region of FWHM no larger than $240$ pc. The authors then use reported line widths of polycyclic aromatic hydrocarbon (PAH) emission -- a known indicator of star-formation activity -- in starburst regions to find that starburst activity can only contribute up to 0.01--0.1 of the total mid-infrared emission, concluding that it must originate instead from a dusty circumnuclear AGN torus. \cite{2009MNRAS.397..265S} have also detected PG~1115+080 in the infrared, using the Infrared Astronomical Satellite (IRAS) to find values close to 1 Jy each at 80 and 100~\textmu m.

More recently, \cite{ae7b5bac19b64fffa43b0dcb87037cbf} used {\it Herschel}/SPIRE to measure emission in the far-IR. The authors used the 250 \textmu m, 350 \textmu m and 500 \textmu m flux densities in addition to the 353~GHz value from \cite{2002ApJ...571..712B} to model a single heated dust component with a modified blackbody spectrum, assuming optically thin dust, obtaining an effective dust temperature $T_e= 52$K. Observations using the Atacama Large (sub)Millimeter Array (ALMA)  have been used to produce a map of  346~GHz emission \citep{10.1093/mnras/staa3433}, which shows a detection of all four lensed components.  The total 346~GHz flux density was used to refine the dust component model, obtaining a fit to the data of $T_e= 66^{+17}_{-21}$~K. The authors used the visibilty data to perform non-parametric modelling of the source  with a singular power law ellipsoid lens plus external shear representing the lensing mass, predicting a source size of effective radius $R_e = 140\pm40$~pc.  While this is consistent, within the errors, with the  source size determined by the 5~GHz emission source model, inspection of the respective maps finds that the emission in the ALMA image, produced using a restoring beam of size 0.32 $\times$ 0.21 arcsec, appears to be more compact than the 5~GHz emission.

\subsubsection{Dust properties}

Returning to the SED of the source, we model the contribution to the lensed IR emission from dust by making use of both the mid-IR and far-IR values from the literature, which correspond to a rest wavelength range 20--330 \textmu  m. According to \cite{1983QJRAS..24..267H}, we model dust components as modified blackbody spectra, assuming an optically thin component:

\begin{equation}
    S_{\nu} \propto \frac{\nu^{3+\beta}}{e^{h\nu/kT_e}-1},
\end{equation}

\noindent where $\nu$ is the rest frequency, $\beta$ is the emissivity index, $T_e$ is the effective dust temperature, $h$ is the Planck constant, $k$ is the Boltzmann constant and $c$ is the speed of light. We make the assumption that each modelled component is optically thin. We note the limitations of this simple parameterisation, which does not account for varying sizes of dust grains  \citep{2012MNRAS.424..951H},  and for which the effective dust temperature is known to be degenerate with emissivity index. We construct four models as follows: M$_{\rm s}$, a single dust component with a dust emissivity index fixed at $\beta=1.5$; M$_{\rm s\beta}$,a single dust component with an emissivity index allowed to vary; M$_{\rm d}$, two dust components each with a dust emissivity index fixed at $\beta=1.5$; M$_{\rm d\beta}$, two dust components each with an emissivity index allowed to vary. In all cases we allow the normalisation constant $A$ to vary for each component. We use least squares (LS) fitting to perform initial model selection. In each case, the numbers of constraints and free parameters are used in conjunction with the minimised value of the objective function to obtain a reduced chi-squared value and a Bayesian inference criterion (BIC) value. Table~\ref{model_select} reports parameter values and goodness-of-fit measurements for each model. From the reduced chi-squared values the four models are ranked in order of preference to obtain two favoured models: M$_{\rm s\beta}$ and M$_{\rm d}$. Approximating the weight of evidence, log $B_{12}$, using the BIC values:

\begin{equation}
    2 \:{\rm log }B_{12} = -{\rm BIC_1}+{\rm BIC_2},
\end{equation}

we determine a Bayes factor $B_{12}$ of 1:1.9 in favour of M$_{\rm s\beta}$, over M$_{\rm d}$. Given that this is not a strong preference, we retain both models for consideration.

\begin{figure}

      \includegraphics[trim={0cm 0cm 1cm 1cm},clip,width=1\linewidth]{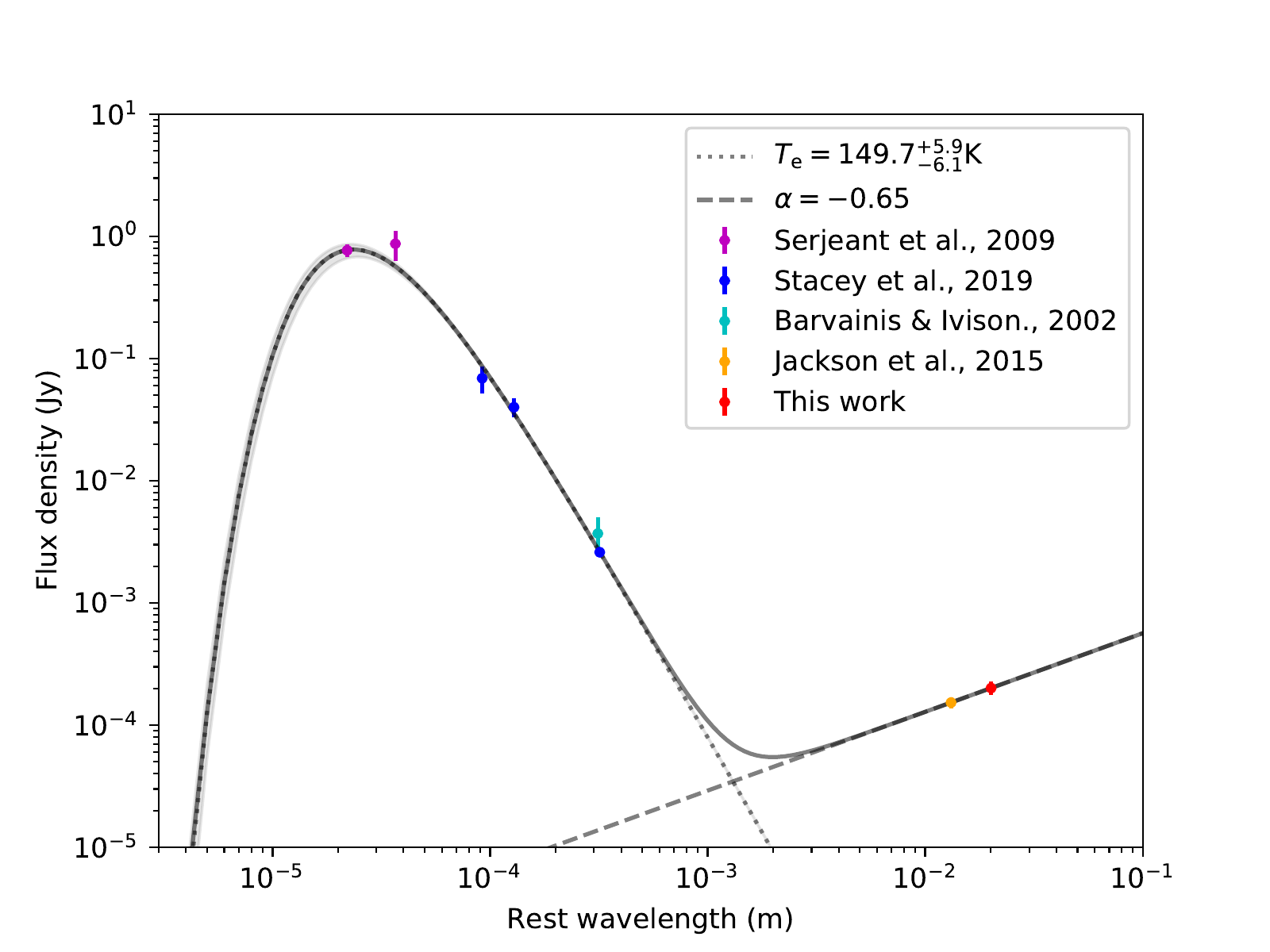}
  \caption{Millimetre to centimetre SED (solid curve) plotted using values from the literature. A warm dust component (dotted curve) is modelled as a modified blackbody spectrum and a non-thermal component (dashed curve) is parameterised as a power law. The grey shaded region represents the range of SEDs produced between the 16th and 84th percentile values of the warm component model parameters.} 

 \label{1115sedone}
 
 \end{figure}

\begin{figure}

      \includegraphics[trim={0cm 0cm 1cm 1cm},clip,width=1\linewidth]{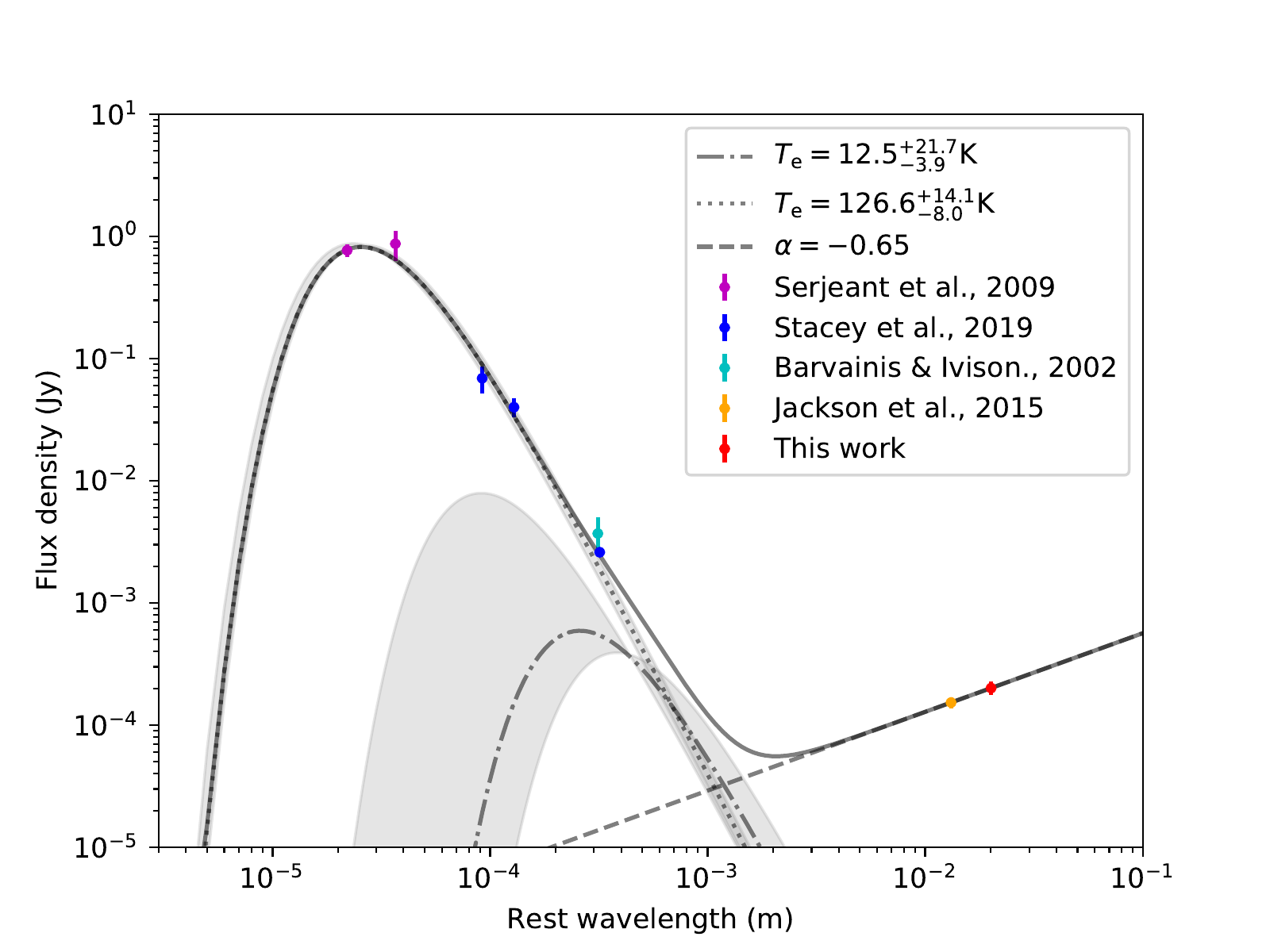} 

  \caption{Millimetre to centimetre SED (solid curve) plotted using values from the literature. Two dust components -- cool (dot-dashed curve) and warm (dotted curve) -- are modelled as modified blackbody spectra and a non-thermal component (dashed curve) is parameterised as a powerlaw. The grey shaded region represents the range of SEDs produced between the 16th and 84th percentile values of the model parameters for each component.} 

 \label{1115sed}
 
 \end{figure}

Performing MCMC sampling in order to obtain the posterior distribution for model parameters, we find that the single component model predicts a warm dust component of effective temperature $T_{e,\rm w} =149.7^{+5.9}_{-6.1}$K and emissivity index $\beta = 1.15^{+0.17}_{-0.07}$ (Fig.~\ref{1115sedone}). The double component model predicts a cold dust component of $T_{e,\rm c}=12.5^{+21.7}_{-3.9}$K and a warm component of $T_{e,\rm w} =126.6^{+14.1}_{-8.0}$K (Fig.~\ref{1115sed}). Our results therefore define two alternative scenarios that could explain the SED: one scenario featuring a a single, warm, dust component, and one featuring a warm and a cool component. New data points could distinguish between these scenarios, could allow the temperatures to be constrained more tightly, and could also provide constraints on the emissivity index in the two component model.

\begin{table*}
	\centering
		{\footnotesize
	\begin{tabular}{lccccccccccc} 
	
		   	\hline
        	\hline\noalign{\smallskip}
    	  Model&	 $T_{e,\rm w}$ (K)&$T_{e,\rm c}$ (K)&$\beta_{\rm w}$&$\beta_{\rm c}$ &$L_{\rm FIR, c} (L_{\odot})$ & $L_{\rm FIR, w} (L_{\odot})$ &$q_{\rm IR,w}$&$q_{\rm IR,c}$ & $\chi^2_{\rm red}$ & BIC \\

    	 	 \hline\noalign{\smallskip}
			   M$_{\rm s}$  &117&-&$\bf 1.5$&-&$2.27\times 10^{13}$&-&2.83&-&2.22&12.5\\
			   M$_{\rm s\beta}$&150&-&1.14&-&$1.70\times 10^{13}$&-&2.71&-&1.18&8.93\\
			  M$_{\rm d}$ &131&31.3&$\bf1.5$&$\bf 1.5$&$1.68\times 10^{13}$ &$4.23\times 10^{11}$&2.70&1.10&1.57&10.3\\
			   M$_{\rm d\beta}$ &123&32.8&1.69&1.50&$1.68  \times 10^{13}$&$6.73  \times 10^{11}$&2.70&1.31&$\inf$&13.71\\

		\noalign{\smallskip}\hline
		  	\hline \noalign{\smallskip}
	\end{tabular}
        }
    	\caption{Model selection performed by optimising using least squares. Models containing single (s) and double(d) dust components are considered, in addition to models with and without a freely-varying emissivity index ($\beta$). In each case we report the best-fit model parameters -- with fixed parameters indicated in bold --  along with the goodness-of-fit metrics.}
        \label{model_select}
\end{table*}

In order to attribute physical origins to the modelled dust components, we compare the modelled effective temperatures with values from the literature. The temperature of the warm dust component in both models is warmer than observations would suggest is possible for star-related activity. Radiative transfer models of clumpy circumnuclear tori predict emission that peaks in the near-infrared, with temperatures above 1000~K \citep{1987ApJ...320..537B,2006MNRAS.367L..57R}. On the cooler side, direct observations of galaxy NGC1068 have found a 320~K toroidal dust structure surrounding a smaller hotter structure \citep{2004Natur.429...47J}, while effective temperatures of $\sim$200--300K have been predicted by \cite{2010A&A...523A..27H} and observed by \cite{2011A&A...536A..78K} for optically thick torus dust clouds further from the central AGN.  A torus model alone would not therefore appear to explain the $\sim20-40$\,\textmu m peak in the SED of PG~1115+080.  Instead, it is possible that a second AGN-related feature is responsible for the warm dust emission. 

Recently, \cite{10.1093/mnras/stw667} have used a sample of $z<0.18$ unobscured and optically luminous quasars from the Palomar Green survey to obtain intrinsic spectral energy distributions for AGN. The authors isolate the contribution to the SED from AGN by using the 11.3\textmu m PAH feature to remove the contribution from stars. The results find a far-IR excess with respect to torus-only AGN models, which the authors suggest is likely to stem from AGN-heated dust in the host galaxy at kpc scales and, consequently, that FIR emission cannot safely be used to calculate star formation rates without subtracting the AGN contribution. \cite{2021arXiv210312747M} also show that AGN can contribute to FIR emission, using radiative transfer models and taking the differences of SEDs to conclude that AGN heating of host-galaxy-scale diffuse dust may contribute to the FIR emission, and again the consequence that star formation rates calculated from the FIR luminosity assuming no AGN contribution can overestimate the true value. \cite{2015ApJ...814....9K} also notice in a sample of 343 (Ultra) Luminous Far Infrared Galaxies that the far-IR emission becomes flatter due to an increase in the warm dust emission. These results have been challenged by \cite{2020ApJ...894...21X} and \cite{2021MNRAS.503.2598B} who, using SED templates featuring dust clouds that are distributed in a torus-like geometry with no contribution from AGN-heated host galaxy dust,  demonstrate a significantly smaller AGN contribution to the FIR than found by \cite{10.1093/mnras/stw667} and \cite{2021arXiv210312747M}.

In the double dust component model, the temperature of the cold component is consistent with those of regions containing old stars and also star-forming regions. \cite{2020MNRAS.499.5732K}, for example, have combined radiation hydrodynamics with non-equilibrium thermochemistry to simulate the interstellar medium (ISM). The simulations predict dust temperatures of $\sim$18 K for regions heated only by old stars, and $\sim$30-40~K for star-forming regions, in agreement with the dust temperatures of star forming regions of local group galaxies \citep{Rela_o_2009,Tabatabaei19,Utomo_2019}.  Further, \cite{ae7b5bac19b64fffa43b0dcb87037cbf} selected a sample of high redshift ($z\sim$2)  dusty star forming galaxies from the study of \cite{2012A&A...539A.155M} and excluded all objects with evidence of a strong AGN component to find a median effective dust temperature of 38$^{+12}_{-5}$~K for 46 objects. The lack of a cold component in the single dust component model points to minimal star-formation activity.

\subsubsection{Comparison with the radio--FIR correlation}

In order to investigate the origin of the radio emission in this source we return to the 5~GHz VLA observations and compare the radio luminosity with the modelled FIR luminosity. In the single dust component model, negligible contribution from star-forming activity points to radio emission that originates from the AGN engine. In the double dust component model, star-forming activity giving rise to the cold dust component would also result in radio emission. A correlation between FIR and radio luminosities has long been established for star-forming galaxies and star-burst regions \citep{1985A&A...147L...6D,1985ApJ...298L...7H,1991ApJ...376...95C}, and is attributed to the connection between young massive stars and their end products. The correlation is quantified by the ratio $q_{\rm IR}$ of rest-frame 8--1000-\textmu m flux, $L_{\rm IR}$, to monochromatic radio luminosity $L_{1.4\;\rm GHz}$:

\begin{equation}
    q_{\rm IR} = {\rm log}_{10}\left( \frac{L_{\rm IR}}{3.75\times 10^{12}L_{1.4\; \rm GHz}} \right),
\end{equation}

\noindent which has been determined using {\it Herschel} and VLA observations of the GOODS North field to be $q_{\rm IR} = 2.40\pm0.24$  \citep{2010A&A...518L..31I}. Evolution of the correlation with redshift $z$ has been observed in some studies. For example,  \cite{2015A&A...573A..45M} find the relation   $q_{\rm IR}(z) = (2.35\pm 0.08)(1+z)^{(-0.12\pm 0.04)}+\log_{10}(1.91)$ and  \cite{2017A&A...602A...4D} find $q_{\rm IR}(z) = (2.88\pm 0.03)(1+z)^{(-0.19\pm 0.01)}$, producing the ratios $q_{\rm IR} = 2.37\pm 0.12$ and $q_{\rm IR} = 2.38\pm 0.04$ at the redshift of PG~1115+080. The correlation has also been observed in the radio-quiet quasar population   \citep{doi:10.1093/mnras/251.1.14P}, and has, therefore,  traditionally been used to rule out AGN activity as the source of radio emission in those quasars where no radio emission in excess of the correlation is found.

In order to investigate whether the star-formation of the double dust component model could fully explain the radio emission of PG~1115+080,  we calculate the $q_{\rm IR }$ ratio using the radio luminosity of the lensed source components with the lensed IR luminosity from the cold dust component only, having removed the contribution from the non-star-forming warm dust component. We first calculate the lensed FIR luminosity $L_{\rm FIR}$ of the component by integrating the FIR emission of the fitted modified blackbody component over rest wavelengths 40 to 120 \textmu m:

\begin{equation}
    L_{\rm FIR} = \frac{4\pi D_{\rm L}^2}{1+z}  \int_{120 \; \rm \mu m}^{40 \; \rm \mu m}   S_{\rm \nu,rest}\;\rm d_{\nu},
\end{equation}

\noindent where $D_{\rm L}$ is the luminosity distance. We obtain, in units of solar luminosity L$_\odot$, $L_{\rm FIR,c}=2.26^{+405}_{-2.21}\times 10^{9}$L$_\odot$. We use the colour correction factor of 1.91 from \cite{2001ApJ...549..215D} in order to extrapolate to the wavelength range of 8--1000~\textmu m via $L_{\rm IR} = 1.91 L_{\rm FIR}$. The IR luminosity allows us to calculate a rate of star-formation associated with the cold dust component in the two component model. For this we use the conversion from \cite{Kennicutt_Jr__1998},

\begin{equation}
{\rm SFR \;(M}_\odot \;{\rm yr}^{-1}) = \frac{L_{\rm IR}}{5.8\times 10^9},
\end{equation}    
    
\noindent assuming a Salpeter initial mass function, to determine a maximum rate of star-formation SFR  $9.58$~M$_\odot$ yr$^{-1}$, having used the FIR magnification factor of 21.5 from \cite{10.1093/mnras/staa3433} to arrive at the unlensed IR luminosity in units of solar luminosity. 

The total lens-magnified 5 GHz flux density of the source, at $179\pm22$ \textmu Jy, corresponds to a K-corrected luminosity at $L_{\rm 1.4 GHz}$ of $6.4\pm3.5\times 10^{24}$ W Hz$^{-1}$, making use of the spectral index of $\alpha = -0.65\pm0.38$. We therefore find for the cold dust component a ratio $q_{\rm IR,c} =-1.16^{+2.25}_{-1.16}$. While not well constrained using the available data, this value falls below the value  found by $q_{\rm IR} = 2.40\pm0.24$, and by  \cite{2015A&A...573A..45M}  and  \cite{2017A&A...602A...4D} for star-forming regions at the redshift of the source. We tentatively suggest that the $q_{\rm IR}$ value is inconsistent with the radio--FIR correlation for star-forming regions and that most -- or all, in the case of the single dust component model -- of the radio emission we observe at 5~GHz originates not from star-forming activity but from an AGN mechanism.

Given that the SED of PG~1115+080 appears to include a warm dust feature resulting neither from the AGN torus nor from star-forming activity, and given the suggestion in the literature that such a component may result from AGN-heated dust in the host galaxy, the result leads us to hypothesize that the warm dust emission may be directly connected to the origin of the radio emission, perhaps from a small jet which is able to heat host-galaxy dust close to the central AGN engine via recycled UV radiation produced as jet shock fronts collide with interstellar gas \citep{10.1046/j.1365-8711.2001.04916.x}. We use 5~GHz luminosity and the IR luminosity of the warm component in both the single and double dust component models to calculate their position with respect to the radio--FIR correlation. For the single component model, we find $L_{\rm FIR,w}=1.69^{+0.07}_{-0.07}\times 10^{13}$L$_\odot$ and $q_{\rm IR,w} =2.70^{+0.02}_{-0.2}$. For the double dust component model, we find $L_{\rm FIR,w}=1.84^{+0.29}_{-0.43}\times 10^{13}$L$_\odot$ and $q_{\rm IR,w} =2.60^{+0.06}_{-0.12}$. These values are consistent with the radio--FIR correlation observed for star-forming regions, despite the effective temperature of the component being above observed values of star-forming or star-burst regions. This suggests, therefore, that the radio--FIR correlation may not be a reliable way of distinguishing between AGN and star-formation related mechanisms of radio emission in radio quiet quasars. Instead, the correlation could conceal additional radio- and FIR-producing mechanisms, such as those due to AGN activity. The result also supports the suggestions of \cite{10.1093/mnras/stw667}, \cite{2021arXiv210312747M} and \cite{2015ApJ...814....9K} that star-formation rates may be over-estimated when assuming no AGN heating of host-galaxy-scale dust.

Strong evidence for sub-pc twin jets spanning 40 pc in RQQ HS~0810+2554 has previously been found \citep{har19}, in a source which sits within the scatter of the radio--FIR correlation when modelled using cold and warm dust components. Both HS~0810+2554 and PG~1115+080 have been found to contain relativistic X-ray absorbing outflows, with inferred velocities of outflowing components ranging between $\sim$0.1c and $\sim$0.4c in each \citep{0004-637X-595-1-85,2007AJ....133.1849C,2014ApJ...783...57C,2016ApJ...824...53C}. We therefore make a very tentative hypothesis that mini jets have coupled with the X-ray-absorbing gas to produce the outflows.   We note that emission originating from two individual dust components may experience differential lensing magnification. The effect of this could be to boost the relative contribution from either star-forming or AGN activity in both the IR and the radio and could the bias $q_{\rm IR}$ value, which may require a revision accounting for differential macro-magnification. The very high resolution afforded by VLBI observations would again be able to confirm AGN-originating radio emission in this source.

\section{Discussion and conclusions}

\begin{table*}
	\centering
    
	\begin{tabular}{llllll}
	    \hline 
        \hline \noalign{\smallskip}
	   System& $z_{\rm lens}$&$z_{\rm source}$&Radio emission & Evidence & References\\
	        
	   & && mechanism &  & \\
       		\hline \noalign{\smallskip}
	   HS~0810+2554 & $\sim$0.8&1.51&AGN dominant* & Twin compact radio components&\cite{har19} \\
       PG~1115+080 &0.31  &1.72&AGN dominant & Dust components; q$_{\rm IR}$ & This work\\ 
       RXJ~1131-1231 &0.30  &0.66&SF  & No compact components; turbulent gas&\cite{2008arXiv0811.3421W}; \cite{paraficz2018alma}  \\
       SDSS~J0924+0219 &0.39  &1.52&SF & Radio coincident with rotating disk & \cite{Badole_2020}\\
       HE 0435-1223 & 0.45 &1.69 &AGN/SF  & Moderately extended source & \cite{2015MNRAS.454..287J}\\
            SSDS J1004+4112&0.68 &1.73& AGN dominant & Intrinsic variability of the radio source& This work; \cite{2021MNRAS.tmpL..33M} \\
                  RX~J0911+0551&0.77 &2.76&AGN/SF & Moderately extended source & \cite{2015MNRAS.454..287J}  \\
                 
		\noalign{\smallskip}\hline
        \hline
	\end{tabular}
    	\caption{Summary of systems which have been imaged with high ($<$1 arcsec) resolution in the radio to determine whether the dominant emission mechanism in the radio is star formation (SF) or AGN activity.  The latter is typically determined by the discovery of small, extended sources; an asterisk denotes confirmation using VLBI. Both emission mechanisms can occur in any one source. See \citet{2018MNRAS.476.5075S} for evidence for star-formation in the general population of radio-quiet quasars using IR-to-radio SEDs.}
        \label{systems}
\end{table*}

We have presented the results from 5 GHz VLA observations of two quadruply-lensed RQQs. Observations of these objects are important, because the magnification afforded by strong lensing allows us direct access to very faint radio source populations out to cosmic noon, allowing us to make maps of the lensed source radio emission that would not otherwise be possible. With high resolution radio maps in particular, we are able to place constraints on source structure that enable us to determine the nature of the emission and begin to resolve the longstanding problem of radio emission in radio-quiet quasars. Understanding whether RQQ radio emission results from star-formation, from some form of AGN activity, or from both, is crucial if we are to understand galaxy feedback models, their role in star-formation `quenching', and galaxy evolution as a whole. 

In our observations of SDSS~J1004+4112 and PG~1115+080 we have found strong and tentative evidence, respectively, for AGN jet activity. In SDSS~J1004+4112, we find an apparent variability of the source in the radio that co-occurs with a similar variability in the optical. We rule out all explanations for variability that do not require the radio source to be of AGN origin. In PG~1115+080, we use photometric infrared flux measurements from the literature which allow us to model contributions to the SED from warm and cool dust components. We find two plausible models: one described fully by a warm dust component only, and one with contributions from warm and cool dust components. By removing the warm component from both models -- likely arising from AGN activity -- we find evidence either for minimal star-formation activity or for an excess of radio emission with respect to the radio--FIR correlation observed for star-forming regions. We therefore make a tentative conclusion that at least some of the radio emission originates from AGN activity. 

Our recent observations bring to a total of 7 the number of quadruply-lensed RQQs that have undergone high (<1 \arcsec ) resolution radio observations (see Table~\ref{systems}). The sample, which uses the magnification properties of strong gravitational lenses to make direct observations of the quasar radio structure, reveals a mixed picture. From this modest number of sources, we see evidence for both star-formation and AGN activity as the dominant radio emission mechanisms within RQQs. These findings are consistent with the contradicting nature of results found from statistical investigations, and may explain why debate over RQQ radio emission has been able to persist in the literature for so long. We note the relatively low intrinsic rest-frame radio luminosities of HS~0810+2554 and SDSS~J1004+4112, at  $L_{\rm 5 GHz}\sim2.4\times 10^{22}$ W\,Hz$^{-1}$ and $L_{\rm 5 GHz}\sim3\times 10^{21}$ W\,Hz$^{-1}$, respectively. Both objects show strong evidence for AGN-dominated radio activity, yet both are the among the faintest known radio sources ever to be imaged (see also \cite{2021ApJ...910..105H}), with peak surface brightness values of just $\sim$0.9 and $\sim$2 \textmu Jy/beam, respectively . Both sources are located within the upturn at the faint end of the radio quasar luminosity function, which has previously been modelled using a star-forming population (e.g. \citealt{Kimball_2011}).

Rather than a question of the physical differences between RQQ and RLQ populations, we may instead need to ask what is responsible for the difference within the RQQ population itself. Recent work by \cite{paraficz2018alma}, who studied lensed RQQ RXJ~1131$-$1231 in the far-IR, found that a $\sim$21~kpc rotating disk of molecular gas coincident with a spiral host clearly visible in the optical \citep{2014ApJ...788L..35S,10.1093/mnras/stz2547}, and which appears to be embedded with turbulent regions of ongoing star formation, would explain the radio emission in this case.  Similarities with local and high redshift disk galaxies lead the authors to suggest that the star-forming properties of this and similar systems is dependent on host gas morphology.  The results lead to the tentative idea that the radio emission mechanism of RQQ is not only determined by the activity in the core, but on the morphology of the host galaxy. This idea is supported by the near-IR studies of RLQs, RQQs and radio galaxies (RGs) made by \cite{10.1093/mnras/283.3.930}, who found that while RLQs and RGs have a de Vaucouleurs host galaxy morphology that is shared by half of the RQQ population, the other half of RQQs have an exponential disk morphology. Indeed, VLA and ALMA observations of SDSS J0924+0219 found evidence for a CO molecular disk coincident in size and orientation with radio emission, suggesting star-formation is responsible for the radio emission in this source \citep{Badole_2020}.

Meanwhile, in RQQs with evident AGN-driven radio emission, authors have begun to use observational evidence to understand the role small-scale jets may play in the so-called 'quasar' feedback mode of RQQs. \cite{Jarvis_2019} used radio morphology to favour the existence of low-powered jets in sources known to have ionised outflows, and  \cite{10.1093/mnras/stx2209} infer the presence of jet activity as the cause of feedback 'bubbles' in a RQQ. Hypotheses have been constructed which designate the jet as the driving mechanism behind the wind or, conversely, the wind as a source collimation for the jet (see e.g. \citealt{2021MNRAS.502.4154R}). The findings, showing the coexistence of two forms of quasar feedback, are consistent with multiwavelength observations of lensed objects HS~0810+2554 and PG~1115+080;  both are RQQs of very faint radio emission which show evidence for small-scale jet activity in addition both to wider-angled wind outflows (\citealt{2014ApJ...783...57C,2016ApJ...824...53C} and \citealt{0004-637X-595-1-85,2007AJ....133.1849C}, respectively) and to absorption lines close to the accretion disk (narrow and broad, respectively).  These findings are also consistent with those of \cite{Kratzer_2015}, who find an anti-correlation between wind speeds and the radio loud fraction in the quasar population. Perhaps, in those quasars able to produce jets, the feedback mode is determined by whether the jet is powerful enough to drill through the host galaxy atmosphere -- heating it in the process to prevent star-formation -- or is not powerful enough to emerge from the host galaxy atmosphere, instead coupling to it and clearing star-forming material from the centre of the galaxy \citep{morgantiraf}. 

The total radio and infrared luminosities of objects PG~1115+080 and HS~0810+2554 appear to place them on the radio--FIR correlation, commonly used to rule out AGN activity as a source of radio emission when no radio excess is seen over the infrared. Indeed, \cite{2016MNRAS.455.4191Z} have found AGN-dominated radio activity in a sample of the radio-quiet majority of 160 obscured and unobscured quasars which lie close to the correlation.  In PG~1115+080, our 5~GHz observations provide tentative evidence that the position of RQQs on the correlation could in some cases be explained by processes other than star-formation.  In AGN-dominated RQQs, it may be that small-scaled jets are able to heat host galaxy dust via shock front interaction, in a feedback mode that conspires to place the object on the radio--FIR correlation.   Sub-millimetre observations of SDSS~J1004+4112 and other RQQs where AGN activity is responsible for the faint radio emission can further test this hypothesis. We suggest that these results reinforce the need to use caution when applying the radio--FIR correlation to attribute radio emission to star-formation alone and when using FIR emission to estimate star-formation rates. 

Our recent observations also provide a further two sky brightness distribution maps that can be used for the study of intervening dark matter substructure.  The resolution of the VLA observations in A configuration has allowed us to rule out the presence of  $>2.5\times 10^6\;{\rm M}_{\odot}$ dark matter structures in the vicinity of the merging components of PG~1115+080, which display a modest flux ratio anomaly in the radio that can be explained by an extended source structure. This finding is consistent with those of \citet{2010MNRAS.406.2764T}, who ascribe a flux ratio in the optical to microlensing. Our observations of SDSS J1004+4112 reveal an anomalous flux density value at component C when compared with the smooth mass models of \citet{oguri2010mass}. The anomaly could be the result of substructures in the galaxy or along the line of sight, but could also be explained by the intrinsic variability of the source that is evident in the variation of component D. Further observations can determine whether the anomaly remains over an extended time and thus allow the intrinsic variability of the source to be decoupled from any milli - or microlensing effects, providing a window on the cold dark matter paradigm. In both objects, the greater understanding of the source structure will reduce the number of assumptions that need to be made when modelling the intervening matter. Future dedicated studies will investigate the mass structure of these objects in more detail.

\section*{Acknowledgements}

The VLA is operated by the US National Radio Astronomy Observatory (NRAO). The NRAO is a facility of the National Science Foundation operated under cooperative agreement by Associated Universities, Inc. This work received support from the European Research Council (ERC) under the European Union's Horizon 2020 research and innovation program (COSMICLENS: grant agreement No 787886). JPM acknowledges support from the Netherlands Organization for Scientific Research (NWO) (Project No. 629.001.023) and the Chinese Academy of Sciences (CAS) (Project No. 114A11KYSB20170054). We thank the anonymous referee for helpful comments on the paper. 

\section*{Data availability}
The radio data used in this work has been obtained by the VLA. The uncalibrated dataset is public and available from the VLA data archive (project 16B-151).




\bibliographystyle{mnras}
\bibliography{1004,phd}

\begin{thebibliography}{}
\makeatletter
\relax
\def\mn@urlcharsother{\let\do\@makeother \do\$\do\&\do\#\do\^\do\_\do\%\do\~}
\def\mn@doi{\begingroup\mn@urlcharsother \@ifnextchar [ {\mn@doi@}
  {\mn@doi@[]}}
\def\mn@doi@[#1]#2{\def\@tempa{#1}\ifx\@tempa\@empty \href
  {http://dx.doi.org/#2} {doi:#2}\else \href {http://dx.doi.org/#2} {#1}\fi
  \endgroup}
\def\mn@eprint#1#2{\mn@eprint@#1:#2::\@nil}
\def\mn@eprint@arXiv#1{\href {http://arxiv.org/abs/#1} {{\tt arXiv:#1}}}
\def\mn@eprint@dblp#1{\href {http://dblp.uni-trier.de/rec/bibtex/#1.xml}
  {dblp:#1}}
\def\mn@eprint@#1:#2:#3:#4\@nil{\def\@tempa {#1}\def\@tempb {#2}\def\@tempc
  {#3}\ifx \@tempc \@empty \let \@tempc \@tempb \let \@tempb \@tempa \fi \ifx
  \@tempb \@empty \def\@tempb {arXiv}\fi \@ifundefined
  {mn@eprint@\@tempb}{\@tempb:\@tempc}{\expandafter \expandafter \csname
  mn@eprint@\@tempb\endcsname \expandafter{\@tempc}}}

\bibitem[\protect\citeauthoryear{Argo, Muxlow, Pedlar, Beswick  \& Strong}{Argo
  et~al.}{2004}]{10.1111/j.1365-2966.2004.08004.x}
Argo M.~K.,  Muxlow T. W.~B.,  Pedlar A.,  Beswick R.~J.,   Strong M.,  2004,
  \mn@doi [Monthly Notices of the Royal Astronomical Society]
  {10.1111/j.1365-2966.2004.08004.x}, 351, L66

\bibitem[\protect\citeauthoryear{Badole, Jackson, Hartley, Sluse, Stacey  \&
  Vives-Arias}{Badole et~al.}{2020}]{Badole_2020}
Badole S.,  Jackson N.,  Hartley P.,  Sluse D.,  Stacey H.,   Vives-Arias H.,
  2020, \mn@doi [Monthly Notices of the Royal Astronomical Society]
  {10.1093/mnras/staa1488}, 496, 138–151

\bibitem[\protect\citeauthoryear{{Bartelmann}}{{Bartelmann}}{2010}]{2010CQGra..27w3001B}
{Bartelmann} M.,  2010, \mn@doi [Classical and Quantum Gravity]
  {10.1088/0264-9381/27/23/233001}, \href
  {https://ui.adsabs.harvard.edu/abs/2010CQGra..27w3001B} {27, 233001}

\bibitem[\protect\citeauthoryear{{Barvainis}}{{Barvainis}}{1987}]{1987ApJ...320..537B}
{Barvainis} R.,  1987, \mn@doi [\apj] {10.1086/165571}, \href
  {https://ui.adsabs.harvard.edu/abs/1987ApJ...320..537B} {320, 537}

\bibitem[\protect\citeauthoryear{{Barvainis} \& {Ivison}}{{Barvainis} \&
  {Ivison}}{2002}]{2002ApJ...571..712B}
{Barvainis} R.,  {Ivison} R.,  2002, \mn@doi [\apj] {10.1086/340096}, \href
  {https://ui.adsabs.harvard.edu/abs/2002ApJ...571..712B} {571, 712}

\bibitem[\protect\citeauthoryear{{Barvainis}, {Leh{\'a}r}, {Birkinshaw},
  {Falcke}  \& {Blundell}}{{Barvainis} et~al.}{2005}]{2005ApJ...618..108B}
{Barvainis} R.,  {Leh{\'a}r} J.,  {Birkinshaw} M.,  {Falcke} H.,   {Blundell}
  K.~M.,  2005, \mn@doi [\apj] {10.1086/425859}, \href
  {https://ui.adsabs.harvard.edu/abs/2005ApJ...618..108B} {618, 108}

\bibitem[\protect\citeauthoryear{{Bernhard}, {Tadhunter}, {Mullaney},
  {Grimmett}, {Rosario}  \& {Alexander}}{{Bernhard}
  et~al.}{2021}]{2021MNRAS.503.2598B}
{Bernhard} E.,  {Tadhunter} C.,  {Mullaney} J.~R.,  {Grimmett} L.~P.,
  {Rosario} D.~J.,   {Alexander} D.~M.,  2021, \mn@doi [\mnras]
  {10.1093/mnras/stab419}, \href
  {https://ui.adsabs.harvard.edu/abs/2021MNRAS.503.2598B} {503, 2598}

\bibitem[\protect\citeauthoryear{{Best} \& {Heckman}}{{Best} \&
  {Heckman}}{2012}]{2012MNRAS.421.1569B}
{Best} P.~N.,  {Heckman} T.~M.,  2012, \mn@doi [\mnras]
  {10.1111/j.1365-2966.2012.20414.x}, \href
  {https://ui.adsabs.harvard.edu/abs/2012MNRAS.421.1569B} {421, 1569}

\bibitem[\protect\citeauthoryear{{Biggs}, {Wucknitz}, {Porcas}, {Browne},
  {Jackson}, {Mao}  \& {Wilkinson}}{{Biggs} et~al.}{2003}]{2003MNRAS.338..599B}
{Biggs} A.~D.,  {Wucknitz} O.,  {Porcas} R.~W.,  {Browne} I.~W.~A.,  {Jackson}
  N.~J.,  {Mao} S.,   {Wilkinson} P.~N.,  2003, \mn@doi [\mnras]
  {10.1046/j.1365-8711.2003.06050.x}, \href
  {https://ui.adsabs.harvard.edu/\#abs/2003MNRAS.338..599B} {338, 599}

\bibitem[\protect\citeauthoryear{Blandford, Meier  \& Readhead}{Blandford
  et~al.}{2019}]{Blandford_2019}
Blandford R.,  Meier D.,   Readhead A.,  2019, \mn@doi [Annual Review of
  Astronomy and Astrophysics] {10.1146/annurev-astro-081817-051948}, 57,
  467–509

\bibitem[\protect\citeauthoryear{{Blundell} \& {Beasley}}{{Blundell} \&
  {Beasley}}{1998}]{1998AAS...19311004B}
{Blundell} K.~M.,  {Beasley} A.~J.,  1998, in American Astronomical Society
  Meeting Abstracts. p. 110.04

\bibitem[\protect\citeauthoryear{{Blundell} \& {Kuncic}}{{Blundell} \&
  {Kuncic}}{2007}]{2007ApJ...668L.103B}
{Blundell} K.~M.,  {Kuncic} Z.,  2007, \mn@doi [\apjl] {10.1086/522695}, \href
  {https://ui.adsabs.harvard.edu/abs/2007ApJ...668L.103B} {668, L103}

\bibitem[\protect\citeauthoryear{{Bonzini}, {Padovani}, {Mainieri},
  {Kellermann}, {Miller}, {Rosati}, {Tozzi}  \& {Vattakunnel}}{{Bonzini}
  et~al.}{2013}]{2013MNRAS.436.3759B}
{Bonzini} M.,  {Padovani} P.,  {Mainieri} V.,  {Kellermann} K.~I.,  {Miller}
  N.,  {Rosati} P.,  {Tozzi} P.,   {Vattakunnel} S.,  2013, \mn@doi [\mnras]
  {10.1093/mnras/stt1879}, \href
  {http://adsabs.harvard.edu/abs/2013MNRAS.436.3759B} {436, 3759}

\bibitem[\protect\citeauthoryear{{Bonzini} et~al.,}{{Bonzini}
  et~al.}{2015}]{2015MNRAS.453.1079B}
{Bonzini} M.,  et~al., 2015, \mn@doi [\mnras] {10.1093/mnras/stv1675}, \href
  {http://adsabs.harvard.edu/abs/2015MNRAS.453.1079B} {453, 1079}

\bibitem[\protect\citeauthoryear{{Calistro Rivera} et~al.,}{{Calistro Rivera}
  et~al.}{2017a}]{2017MNRAS.469.3468C}
{Calistro Rivera} G.,  et~al., 2017a, \mn@doi [\mnras] {10.1093/mnras/stx1040},
  \href {https://ui.adsabs.harvard.edu/abs/2017MNRAS.469.3468C} {469, 3468}

\bibitem[\protect\citeauthoryear{Calistro~Rivera et~al.,}{Calistro~Rivera
  et~al.}{2017b}]{10.1093/mnras/stx1040}
Calistro~Rivera G.,  et~al., 2017b, \mn@doi [Monthly Notices of the Royal
  Astronomical Society] {10.1093/mnras/stx1040}, 469, 3468

\bibitem[\protect\citeauthoryear{Chartas, Brandt  \& Gallagher}{Chartas
  et~al.}{2003}]{0004-637X-595-1-85}
Chartas G.,  Brandt W.~N.,   Gallagher S.~C.,  2003, The Astrophysical Journal,
  595, 85

\bibitem[\protect\citeauthoryear{{Chartas}, {Brandt}, {Gallagher}  \&
  {Proga}}{{Chartas} et~al.}{2007}]{2007AJ....133.1849C}
{Chartas} G.,  {Brandt} W.~N.,  {Gallagher} S.~C.,   {Proga} D.,  2007, \mn@doi
  [\aj] {10.1086/512364}, \href
  {https://ui.adsabs.harvard.edu/#abs/2007AJ....133.1849C} {133, 1849}

\bibitem[\protect\citeauthoryear{{Chartas}, {Hamann}, {Eracleous}, {Misawa},
  {Cappi}, {Giustini}, {Charlton}  \& {Marvin}}{{Chartas}
  et~al.}{2014}]{2014ApJ...783...57C}
{Chartas} G.,  {Hamann} F.,  {Eracleous} M.,  {Misawa} T.,  {Cappi} M.,
  {Giustini} M.,  {Charlton} J.~C.,   {Marvin} M.,  2014, \mn@doi [\apj]
  {10.1088/0004-637X/783/1/57}, \href
  {http://adsabs.harvard.edu/abs/2014ApJ...783...57C} {783, 57}

\bibitem[\protect\citeauthoryear{{Chartas}, {Cappi}, {Hamann}, {Eracleous},
  {Strickland}, {Giustini}  \& {Misawa}}{{Chartas}
  et~al.}{2016}]{2016ApJ...824...53C}
{Chartas} G.,  {Cappi} M.,  {Hamann} F.,  {Eracleous} M.,  {Strickland} S.,
  {Giustini} M.,   {Misawa} T.,  2016, \mn@doi [\apj]
  {10.3847/0004-637X/824/1/53}, \href
  {http://adsabs.harvard.edu/abs/2016ApJ...824...53C} {824, 53}

\bibitem[\protect\citeauthoryear{Chen et~al.,}{Chen
  et~al.}{2019}]{10.1093/mnras/stz2547}
Chen G. C.-F.,  et~al., 2019, \mn@doi [Monthly Notices of the Royal
  Astronomical Society] {10.1093/mnras/stz2547}, 490, 1743

\bibitem[\protect\citeauthoryear{{Chiba}}{{Chiba}}{2002}]{2002ApJ...565...17C}
{Chiba} M.,  2002, \mn@doi [\apj] {10.1086/324493}, \href
  {https://ui.adsabs.harvard.edu/abs/2002ApJ...565...17C} {565, 17}

\bibitem[\protect\citeauthoryear{Chiba, Minezaki, Kashikawa, Kataza  \&
  Inoue}{Chiba et~al.}{2005}]{Chiba_2005}
Chiba M.,  Minezaki T.,  Kashikawa N.,  Kataza H.,   Inoue K.~T.,  2005,
  \mn@doi [The Astrophysical Journal] {10.1086/430403}, 627, 53

\bibitem[\protect\citeauthoryear{{Claeskens}, {Sluse}, {Riaud}  \&
  {Surdej}}{{Claeskens} et~al.}{2006}]{2006A&A...451..865C}
{Claeskens} J.-F.,  {Sluse} D.,  {Riaud} P.,   {Surdej} J.,  2006, \mn@doi
  [\aap] {10.1051/0004-6361:20054352}, \href
  {http://adsabs.harvard.edu/abs/2006A%26A...451..865C} {451, 865}

\bibitem[\protect\citeauthoryear{{Condon}, {Anderson}  \& {Helou}}{{Condon}
  et~al.}{1991}]{1991ApJ...376...95C}
{Condon} J.~J.,  {Anderson} M.~L.,   {Helou} G.,  1991, \mn@doi [\apj]
  {10.1086/170258}, \href
  {https://ui.adsabs.harvard.edu/abs/1991ApJ...376...95C} {376, 95}

\bibitem[\protect\citeauthoryear{{Condon}, {Kellermann}, {Kimball},
  {Ivezi{\'c}}  \& {Perley}}{{Condon} et~al.}{2013}]{2013ApJ...768...37C}
{Condon} J.~J.,  {Kellermann} K.~I.,  {Kimball} A.~E.,  {Ivezi{\'c}} v.,
  {Perley} R.~A.,  2013, \mn@doi [\apj] {10.1088/0004-637X/768/1/37}, \href
  {http://adsabs.harvard.edu/abs/2013ApJ...768...37C} {768, 37}

\bibitem[\protect\citeauthoryear{{Courbin}, {Magain}, {Keeton}, {Kochanek},
  {Vanderriest}, {Jaunsen}  \& {Hjorth}}{{Courbin} et~al.}{1997}]{cour}
{Courbin} F.,  {Magain} P.,  {Keeton} C.~R.,  {Kochanek} C.~S.,  {Vanderriest}
  C.,  {Jaunsen} A.~O.,   {Hjorth} J.,  1997, \aap, \href
  {https://ui.adsabs.harvard.edu/\#abs/1997A&A...324L...1C} {324, L1}

\bibitem[\protect\citeauthoryear{{Courbin}, {Saha}  \& {Schechter}}{{Courbin}
  et~al.}{2002}]{2002LNP...608....1C}
{Courbin} F.,  {Saha} P.,   {Schechter} P.~L.,  2002, {Quasar Lensing}.
p.~1

\bibitem[\protect\citeauthoryear{{Croton} et~al.,}{{Croton}
  et~al.}{2006}]{2006MNRAS.365...11C}
{Croton} D.~J.,  et~al., 2006, \mn@doi [\mnras]
  {10.1111/j.1365-2966.2005.09675.x}, \href
  {https://ui.adsabs.harvard.edu/abs/2006MNRAS.365...11C} {365, 11}

\bibitem[\protect\citeauthoryear{Dalal \& Kochanek}{Dalal \&
  Kochanek}{2002}]{Dalal_2002}
Dalal N.,  Kochanek C.~S.,  2002, \mn@doi [The Astrophysical Journal]
  {10.1086/340303}, 572, 25–33

\bibitem[\protect\citeauthoryear{{Dale}, {Helou}, {Contursi}, {Silbermann}  \&
  {Kolhatkar}}{{Dale} et~al.}{2001}]{2001ApJ...549..215D}
{Dale} D.~A.,  {Helou} G.,  {Contursi} A.,  {Silbermann} N.~A.,   {Kolhatkar}
  S.,  2001, \mn@doi [\apj] {10.1086/319077}, \href
  {https://ui.adsabs.harvard.edu/abs/2001ApJ...549..215D} {549, 215}

\bibitem[\protect\citeauthoryear{{Delhaize} et~al.,}{{Delhaize}
  et~al.}{2017}]{2017A&A...602A...4D}
{Delhaize} J.,  et~al., 2017, \mn@doi [\aap] {10.1051/0004-6361/201629430},
  \href {https://ui.adsabs.harvard.edu/abs/2017A&A...602A...4D} {602, A4}

\bibitem[\protect\citeauthoryear{Fian, Mediavilla, Hanslmeier, Oscoz,
  Serra-Ricart, Mu{\~n}oz  \& Jim{\'e}nez-Vicente}{Fian
  et~al.}{2016}]{fian2016size}
Fian C.,  Mediavilla E.,  Hanslmeier A.,  Oscoz A.,  Serra-Ricart M.,
  Mu{\~n}oz J.,   Jim{\'e}nez-Vicente J.,  2016, The Astrophysical Journal,
  830, 149

\bibitem[\protect\citeauthoryear{Fohlmeister et~al.,}{Fohlmeister
  et~al.}{2007}]{Fohlmeister_2007}
Fohlmeister J.,  et~al., 2007, \mn@doi [The Astrophysical Journal]
  {10.1086/518018}, 662, 62

\bibitem[\protect\citeauthoryear{{Fohlmeister}, {Kochanek}, {Falco}, {Morgan}
  \& {Wambsganss}}{{Fohlmeister} et~al.}{2008}]{2008ApJ...676..761F}
{Fohlmeister} J.,  {Kochanek} C.~S.,  {Falco} E.~E.,  {Morgan} C.~W.,
  {Wambsganss} J.,  2008, \mn@doi [\apj] {10.1086/528789}, \href
  {http://adsabs.harvard.edu/abs/2008ApJ...676..761F} {676, 761}

\bibitem[\protect\citeauthoryear{{Foreman-Mackey}, {Hogg}, {Lang}  \&
  {Goodman}}{{Foreman-Mackey} et~al.}{2013}]{2013PASP..125..306F}
{Foreman-Mackey} D.,  {Hogg} D.~W.,  {Lang} D.,   {Goodman} J.,  2013, \mn@doi
  [\pasp] {10.1086/670067}, \href
  {https://ui.adsabs.harvard.edu/abs/2013PASP..125..306F} {125, 306}

\bibitem[\protect\citeauthoryear{{G\"urkan} et~al.,}{{G\"urkan}
  et~al.}{2019}]{gurkanrefId0}
{G\"urkan} G.,  et~al., 2019, \mn@doi [A\&A] {10.1051/0004-6361/201833892},
  622, A11

\bibitem[\protect\citeauthoryear{{Hartley}, {Jackson}, {Sluse}, {Stacey}  \&
  {Vives-Arias}}{{Hartley} et~al.}{2019}]{har19}
{Hartley} P.,  {Jackson} N.,  {Sluse} D.,  {Stacey} H.~R.,   {Vives-Arias} H.,
  2019, \mn@doi [\mnras] {10.1093/mnras/stz510}, \href
  {https://ui.adsabs.harvard.edu/abs/2019MNRAS.485.3009H} {485, 3009}

\bibitem[\protect\citeauthoryear{{Hawkins}}{{Hawkins}}{2020}]{2020A&A...643A..10H}
{Hawkins} M.~R.~S.,  2020, \mn@doi [\aap] {10.1051/0004-6361/202038670}, \href
  {https://ui.adsabs.harvard.edu/abs/2020A&A...643A..10H} {643, A10}

\bibitem[\protect\citeauthoryear{{Hayward}, {Jonsson}, {Kere{\v{s}}},
  {Magnelli}, {Hernquist}  \& {Cox}}{{Hayward}
  et~al.}{2012}]{2012MNRAS.424..951H}
{Hayward} C.~C.,  {Jonsson} P.,  {Kere{\v{s}}} D.,  {Magnelli} B.,  {Hernquist}
  L.,   {Cox} T.~J.,  2012, \mn@doi [\mnras]
  {10.1111/j.1365-2966.2012.21254.x}, \href
  {https://ui.adsabs.harvard.edu/abs/2012MNRAS.424..951H} {424, 951}

\bibitem[\protect\citeauthoryear{{Helou}, {Soifer}  \&
  {Rowan-Robinson}}{{Helou} et~al.}{1985}]{1985ApJ...298L...7H}
{Helou} G.,  {Soifer} B.~T.,   {Rowan-Robinson} M.,  1985, \mn@doi [\apjl]
  {10.1086/184556}, \href
  {https://ui.adsabs.harvard.edu/abs/1985ApJ...298L...7H} {298, L7}

\bibitem[\protect\citeauthoryear{{Herrera Ruiz}, {Middelberg}, {Norris}  \&
  {Maini}}{{Herrera Ruiz} et~al.}{2016}]{2016A&A...589L...2H}
{Herrera Ruiz} N.,  {Middelberg} E.,  {Norris} R.~P.,   {Maini} A.,  2016,
  \mn@doi [\aap] {10.1051/0004-6361/201628302}, \href
  {http://adsabs.harvard.edu/abs/2016A%26A...589L...2H} {589, L2}

\bibitem[\protect\citeauthoryear{{Heywood} et~al.,}{{Heywood}
  et~al.}{2021}]{2021ApJ...910..105H}
{Heywood} I.,  et~al., 2021, \mn@doi [\apj] {10.3847/1538-4357/abdf61}, \href
  {https://ui.adsabs.harvard.edu/abs/2021ApJ...910..105H} {910, 105}

\bibitem[\protect\citeauthoryear{{Hezaveh} et~al.,}{{Hezaveh}
  et~al.}{2013}]{hezaveh13a}
{Hezaveh} Y.~D.,  et~al., 2013, \mn@doi [\apj] {10.1088/0004-637X/767/2/132},
  \href {http://adsabs.harvard.edu/abs/2013ApJ...767..132H} {767, 132}

\bibitem[\protect\citeauthoryear{{Hildebrand}}{{Hildebrand}}{1983}]{1983QJRAS..24..267H}
{Hildebrand} R.~H.,  1983, \qjras, \href
  {https://ui.adsabs.harvard.edu/abs/1983QJRAS..24..267H} {24, 267}

\bibitem[\protect\citeauthoryear{{Hinojosa-Go\~ni}, {Mu\~noz-Tu\~n\'on, C.}  \&
  {M\'endez-Abreu, J.}}{{Hinojosa-Go\~ni} et~al.}{2016}]{HinojosarefId0}
{Hinojosa-Go\~ni} R.,  {Mu\~noz-Tu\~n\'on, C.}  {M\'endez-Abreu, J.} 2016,
  \mn@doi [A\&A] {10.1051/0004-6361/201527066}, 592, A122

\bibitem[\protect\citeauthoryear{{H{\"o}nig} \& {Kishimoto}}{{H{\"o}nig} \&
  {Kishimoto}}{2010}]{2010A&A...523A..27H}
{H{\"o}nig} S.~F.,  {Kishimoto} M.,  2010, \mn@doi [\aap]
  {10.1051/0004-6361/200912676}, \href
  {https://ui.adsabs.harvard.edu/abs/2010A&A...523A..27H} {523, A27}

\bibitem[\protect\citeauthoryear{{Hovatta} et~al.,}{{Hovatta}
  et~al.}{2014}]{2014AJ....147..143H}
{Hovatta} T.,  et~al., 2014, \mn@doi [\aj] {10.1088/0004-6256/147/6/143}, \href
  {https://ui.adsabs.harvard.edu/abs/2014AJ....147..143H} {147, 143}

\bibitem[\protect\citeauthoryear{{Impey}, {Falco}, {Kochanek}, {Leh{\'a}r},
  {McLeod}, {Rix}, {Peng}  \& {Keeton}}{{Impey}
  et~al.}{1998}]{1998ApJ...509..551I}
{Impey} C.~D.,  {Falco} E.~E.,  {Kochanek} C.~S.,  {Leh{\'a}r} J.,  {McLeod}
  B.~A.,  {Rix} H.-W.,  {Peng} C.~Y.,   {Keeton} C.~R.,  1998, \mn@doi [\apj]
  {10.1086/306521}, \href {http://adsabs.harvard.edu/abs/1998ApJ...509..551I}
  {509, 551}

\bibitem[\protect\citeauthoryear{Inada et~al.,}{Inada
  et~al.}{2003}]{inada2003gravitationally}
Inada N.,  et~al., 2003, Nature, 426, 810

\bibitem[\protect\citeauthoryear{{Inada} et~al.,}{{Inada}
  et~al.}{2005}]{2005PASJ...57L...7I}
{Inada} N.,  et~al., 2005, \mn@doi [\pasj] {10.1093/pasj/57.3.L7}, \href
  {https://ui.adsabs.harvard.edu/abs/2005PASJ...57L...7I} {57, L7}

\bibitem[\protect\citeauthoryear{Inada, Oguri, Falco, Broadhurst, Ofek,
  Kochanek, Sharon  \& Smith}{Inada et~al.}{2008}]{inada2008spectroscopic}
Inada N.,  Oguri M.,  Falco E.~E.,  Broadhurst T.~J.,  Ofek E.~O.,  Kochanek
  C.~S.,  Sharon K.,   Smith G.~P.,  2008, Publications of the Astronomical
  Society of Japan, 60, L27

\bibitem[\protect\citeauthoryear{{Ivison} et~al.,}{{Ivison}
  et~al.}{2010}]{2010A&A...518L..31I}
{Ivison} R.~J.,  et~al., 2010, \mn@doi [\aap] {10.1051/0004-6361/201014552},
  \href {https://ui.adsabs.harvard.edu/abs/2010A&A...518L..31I} {518, L31}

\bibitem[\protect\citeauthoryear{{Jackson}}{{Jackson}}{2011}]{2011ApJ...739L..28J}
{Jackson} N.,  2011, \mn@doi [\apjl] {10.1088/2041-8205/739/1/L28}, \href
  {http://adsabs.harvard.edu/abs/2011ApJ...739L..28J} {739, L28}

\bibitem[\protect\citeauthoryear{{Jackson}}{{Jackson}}{2013}]{2013BASI...41...19J}
{Jackson} N.,  2013, Bulletin of the Astronomical Society of India, \href
  {https://ui.adsabs.harvard.edu/abs/2013BASI...41...19J} {41, 19}

\bibitem[\protect\citeauthoryear{{Jackson}, {Tagore}, {Roberts}, {Sluse},
  {Stacey}, {Vives-Arias}, {Wucknitz}  \& {Volino}}{{Jackson}
  et~al.}{2015}]{2015MNRAS.454..287J}
{Jackson} N.,  {Tagore} A.~S.,  {Roberts} C.,  {Sluse} D.,  {Stacey} H.,
  {Vives-Arias} H.,  {Wucknitz} O.,   {Volino} F.,  2015, \mn@doi [\mnras]
  {10.1093/mnras/stv1982}, \href
  {http://adsabs.harvard.edu/abs/2015MNRAS.454..287J} {454, 287}

\bibitem[\protect\citeauthoryear{{Jaffe} et~al.,}{{Jaffe}
  et~al.}{2004}]{2004Natur.429...47J}
{Jaffe} W.,  et~al., 2004, \mn@doi [\nat] {10.1038/nature02531}, \href
  {https://ui.adsabs.harvard.edu/abs/2004Natur.429...47J} {429, 47}

\bibitem[\protect\citeauthoryear{Jarvis et~al.,}{Jarvis
  et~al.}{2019}]{Jarvis_2019}
Jarvis M.~E.,  et~al., 2019, \mn@doi [Monthly Notices of the Royal Astronomical
  Society] {10.1093/mnras/stz556}, 485, 2710–2730

\bibitem[\protect\citeauthoryear{Jorstad \& Marscher}{Jorstad \&
  Marscher}{2016}]{jorstad2016vlba}
Jorstad S.,  Marscher A.,  2016, Galaxies, 4, 47

\bibitem[\protect\citeauthoryear{{Kannan}, {Marinacci}, {Vogelsberger},
  {Sales}, {Torrey}, {Springel}  \& {Hernquist}}{{Kannan}
  et~al.}{2020}]{2020MNRAS.499.5732K}
{Kannan} R.,  {Marinacci} F.,  {Vogelsberger} M.,  {Sales} L.~V.,  {Torrey} P.,
   {Springel} V.,   {Hernquist} L.,  2020, \mn@doi [\mnras]
  {10.1093/mnras/staa3249}, \href
  {https://ui.adsabs.harvard.edu/abs/2020MNRAS.499.5732K} {499, 5732}

\bibitem[\protect\citeauthoryear{{Kellermann}, {Sramek}, {Schmidt}, {Shaffer}
  \& {Green}}{{Kellermann} et~al.}{1989}]{1989AJ.....98.1195K}
{Kellermann} K.~I.,  {Sramek} R.,  {Schmidt} M.,  {Shaffer} D.~B.,   {Green}
  R.,  1989, \mn@doi [\aj] {10.1086/115207}, \href
  {https://ui.adsabs.harvard.edu/abs/1989AJ.....98.1195K} {98, 1195}

\bibitem[\protect\citeauthoryear{{Kennicutt}}{{Kennicutt}}{1998}]{Kennicutt_Jr__1998}
{Kennicutt} R.,  1998, \mn@doi [The Astrophysical Journal] {10.1086/305588},
  498, 541

\bibitem[\protect\citeauthoryear{Kimball, Kellermann, Condon, Ivezi\'{c}  \&
  Perley}{Kimball et~al.}{2011}]{Kimball_2011}
Kimball A.~E.,  Kellermann K.~I.,  Condon J.~J.,  Ivezi\'{c} v.,   Perley
  R.~A.,  2011, \mn@doi [The Astrophysical Journal]
  {10.1088/2041-8205/739/1/l29}, 739, L29

\bibitem[\protect\citeauthoryear{{Kirkpatrick}, {Pope}, {Sajina}, {Roebuck},
  {Yan}, {Armus}, {D{\'\i}az-Santos}  \& {Stierwalt}}{{Kirkpatrick}
  et~al.}{2015}]{2015ApJ...814....9K}
{Kirkpatrick} A.,  {Pope} A.,  {Sajina} A.,  {Roebuck} E.,  {Yan} L.,  {Armus}
  L.,  {D{\'\i}az-Santos} T.,   {Stierwalt} S.,  2015, \mn@doi [\apj]
  {10.1088/0004-637X/814/1/9}, \href
  {https://ui.adsabs.harvard.edu/abs/2015ApJ...814....9K} {814, 9}

\bibitem[\protect\citeauthoryear{{Kishimoto}, {H{\"o}nig}, {Antonucci},
  {Millour}, {Tristram}  \& {Weigelt}}{{Kishimoto}
  et~al.}{2011}]{2011A&A...536A..78K}
{Kishimoto} M.,  {H{\"o}nig} S.~F.,  {Antonucci} R.,  {Millour} F.,  {Tristram}
  K.~R.~W.,   {Weigelt} G.,  2011, \mn@doi [\aap]
  {10.1051/0004-6361/201117367}, \href
  {https://ui.adsabs.harvard.edu/abs/2011A&A...536A..78K} {536, A78}

\bibitem[\protect\citeauthoryear{{Kochanek} \& {Dalal}}{{Kochanek} \&
  {Dalal}}{2004}]{2004ApJ...610...69K}
{Kochanek} C.~S.,  {Dalal} N.,  2004, \mn@doi [\apj] {10.1086/421436}, \href
  {http://adsabs.harvard.edu/abs/2004ApJ...610...69K} {610, 69}

\bibitem[\protect\citeauthoryear{{Koopmans} \& {de Bruyn}}{{Koopmans} \& {de
  Bruyn}}{2000}]{2000A&A...358..793K}
{Koopmans} L.~V.~E.,  {de Bruyn} A.~G.,  2000, \aap, \href
  {http://adsabs.harvard.edu/abs/2000A%26A...358..793K} {358, 793}

\bibitem[\protect\citeauthoryear{Kratzer \& Richards}{Kratzer \&
  Richards}{2015}]{Kratzer_2015}
Kratzer R.~M.,  Richards G.~T.,  2015, \mn@doi [The Astronomical Journal]
  {10.1088/0004-6256/149/2/61}, 149, 61

\bibitem[\protect\citeauthoryear{{Kundic}, {Cohen}, {Blandford}  \&
  {Lubin}}{{Kundic} et~al.}{1997}]{1997AJ....114..507K}
{Kundic} T.,  {Cohen} J.~G.,  {Blandford} R.~D.,   {Lubin} L.~M.,  1997,
  \mn@doi [\aj] {10.1086/118489}, \href
  {http://adsabs.harvard.edu/abs/1997AJ....114..507K} {114, 507}

\bibitem[\protect\citeauthoryear{{Laor}, {Baldi}  \& {Behar}}{{Laor}
  et~al.}{2018}]{2018arXiv181010245L}
{Laor} A.,  {Baldi} R.~D.,   {Behar} E.,  2018, preprint, \href
  {http://adsabs.harvard.edu/abs/2018arXiv181010245L} {} (\mn@eprint {arXiv}
  {1810.10245})

\bibitem[\protect\citeauthoryear{{Leipski}, {Falcke}, {Bennert}  \&
  {H{\"u}ttemeister}}{{Leipski} et~al.}{2006}]{2006A&A...455..161L}
{Leipski} C.,  {Falcke} H.,  {Bennert} N.,   {H{\"u}ttemeister} S.,  2006,
  \mn@doi [\aap] {10.1051/0004-6361:20054311}, \href
  {https://ui.adsabs.harvard.edu/abs/2006A&A...455..161L} {455, 161}

\bibitem[\protect\citeauthoryear{Liesenborgs, De~Rijcke, Dejonghe  \&
  Bekaert}{Liesenborgs et~al.}{2009}]{liesenborgs2009non}
Liesenborgs J.,  De~Rijcke S.,  Dejonghe H.,   Bekaert P.,  2009, Monthly
  Notices of the Royal Astronomical Society, 397, 341

\bibitem[\protect\citeauthoryear{Lister et~al.,}{Lister
  et~al.}{2016}]{lister2016mojave}
Lister M.~L.,  et~al., 2016, The Astronomical Journal, 152, 12

\bibitem[\protect\citeauthoryear{{Macfarlane} et~al.,}{{Macfarlane}
  et~al.}{2021}]{2021MNRAS.506.5888M}
{Macfarlane} C.,  et~al., 2021, \mn@doi [\mnras] {10.1093/mnras/stab1998},
  \href {https://ui.adsabs.harvard.edu/abs/2021MNRAS.506.5888M} {506, 5888}

\bibitem[\protect\citeauthoryear{{Madau} \& {Dickinson}}{{Madau} \&
  {Dickinson}}{2014}]{2014ARA&A..52..415M}
{Madau} P.,  {Dickinson} M.,  2014, \mn@doi [\araa]
  {10.1146/annurev-astro-081811-125615}, \href
  {http://adsabs.harvard.edu/abs/2014ARA%26A..52..415M} {52, 415}

\bibitem[\protect\citeauthoryear{{Magnelli} et~al.,}{{Magnelli}
  et~al.}{2012}]{2012A&A...539A.155M}
{Magnelli} B.,  et~al., 2012, \mn@doi [\aap] {10.1051/0004-6361/201118312},
  \href {https://ui.adsabs.harvard.edu/abs/2012A&A...539A.155M} {539, A155}

\bibitem[\protect\citeauthoryear{{Magnelli} et~al.,}{{Magnelli}
  et~al.}{2015}]{2015A&A...573A..45M}
{Magnelli} B.,  et~al., 2015, \mn@doi [\aap] {10.1051/0004-6361/201424937},
  \href {https://ui.adsabs.harvard.edu/abs/2015A&A...573A..45M} {573, A45}

\bibitem[\protect\citeauthoryear{{Malefahlo}, {Santos}, {Jarvis}, {White}  \&
  {Zwart}}{{Malefahlo} et~al.}{2020}]{2020MNRAS.492.5297M}
{Malefahlo} E.,  {Santos} M.~G.,  {Jarvis} M.~J.,  {White} S.~V.,   {Zwart} J.
  T.~L.,  2020, \mn@doi [\mnras] {10.1093/mnras/staa112}, \href
  {https://ui.adsabs.harvard.edu/abs/2020MNRAS.492.5297M} {492, 5297}

\bibitem[\protect\citeauthoryear{{Mao} \& {Schneider}}{{Mao} \&
  {Schneider}}{1998}]{1998MNRAS.295..587M}
{Mao} S.,  {Schneider} P.,  1998, \mn@doi [\mnras]
  {10.1046/j.1365-8711.1998.01319.x}, \href
  {http://adsabs.harvard.edu/abs/1998MNRAS.295..587M} {295, 587}

\bibitem[\protect\citeauthoryear{{McKean} et~al.,}{{McKean}
  et~al.}{2021}]{2021MNRAS.tmpL..33M}
{McKean} J.~P.,  et~al., 2021, \mn@doi [\mnras] {10.1093/mnrasl/slab033}, \href
  {https://ui.adsabs.harvard.edu/abs/2021MNRAS.tmpL..33M} {}

\bibitem[\protect\citeauthoryear{{McKinney}, {Hayward}, {Rosenthal},
  {Martinez-Galarza}, {Pope}, {Sajina}  \& {Smith}}{{McKinney}
  et~al.}{2021}]{2021arXiv210312747M}
{McKinney} J.,  {Hayward} C.~C.,  {Rosenthal} L.~J.,  {Martinez-Galarza} J.~R.,
   {Pope} A.,  {Sajina} A.,   {Smith} H.~A.,  2021, arXiv e-prints, \href
  {https://ui.adsabs.harvard.edu/abs/2021arXiv210312747M} {p. arXiv:2103.12747}

\bibitem[\protect\citeauthoryear{{Metcalf} \& {Madau}}{{Metcalf} \&
  {Madau}}{2001}]{2001ApJ...563....9M}
{Metcalf} R.~B.,  {Madau} P.,  2001, \mn@doi [\apj] {10.1086/323695}, \href
  {http://adsabs.harvard.edu/abs/2001ApJ...563....9M} {563, 9}

\bibitem[\protect\citeauthoryear{{Meylan}, {Jetzer}, {North}, {Schneider},
  {Kochanek}  \& {Wambsganss}}{{Meylan} et~al.}{2006}]{2006glsw.conf.....M}
{Meylan} G.,  {Jetzer} P.,  {North} P.,  {Schneider} P.,  {Kochanek} C.~S.,
  {Wambsganss} J.,  eds, 2006, {Gravitational Lensing: Strong, Weak and Micro}
  (\mn@eprint {} {astro-ph/0407232})

\bibitem[\protect\citeauthoryear{{Morganti, Raffaella}, {Veilleux, Sylvain},
  {Oosterloo, Tom}, {Teng, Stacy H.}  \& {Rupke, David}}{{Morganti, Raffaella}
  et~al.}{2016}]{morgantiraf}
{Morganti, Raffaella} {Veilleux, Sylvain} {Oosterloo, Tom} {Teng, Stacy H.}
  {Rupke, David} 2016, \mn@doi [A\&A] {10.1051/0004-6361/201628978}, 593, A30

\bibitem[\protect\citeauthoryear{{Murakami} et~al.,}{{Murakami}
  et~al.}{2007}]{2007PASJ...59S.369M}
{Murakami} H.,  et~al., 2007, \mn@doi [\pasj] {10.1093/pasj/59.sp2.S369}, \href
  {https://ui.adsabs.harvard.edu/abs/2007PASJ...59S.369M} {59, S369}

\bibitem[\protect\citeauthoryear{{Nierenberg} et~al.,}{{Nierenberg}
  et~al.}{2017}]{2017MNRAS.471.2224N}
{Nierenberg} A.~M.,  et~al., 2017, \mn@doi [\mnras] {10.1093/mnras/stx1400},
  \href {http://adsabs.harvard.edu/abs/2017MNRAS.471.2224N} {471, 2224}

\bibitem[\protect\citeauthoryear{Oguri}{Oguri}{2010}]{oguri2010mass}
Oguri M.,  2010, Publications of the Astronomical Society of Japan, 62, 1017

\bibitem[\protect\citeauthoryear{{Oguri} et~al.,}{{Oguri}
  et~al.}{2004}]{2004ApJ...605...78O}
{Oguri} M.,  et~al., 2004, \mn@doi [\apj] {10.1086/382221}, \href
  {http://adsabs.harvard.edu/abs/2004ApJ...605...78O} {605, 78}

\bibitem[\protect\citeauthoryear{{Orienti}, {D'Ammando}, {Giroletti},
  {Giovannini}  \& {Panessa}}{{Orienti} et~al.}{2015}]{2015aska.confE..87O}
{Orienti} M.,  {D'Ammando} F.,  {Giroletti} M.,  {Giovannini} G.,   {Panessa}
  F.,  2015, Advancing Astrophysics with the Square Kilometre Array (AASKA14),
  \href {http://adsabs.harvard.edu/abs/2015aska.confE..87O} {p.~87}

\bibitem[\protect\citeauthoryear{{Padovani}}{{Padovani}}{2017}]{2017NatAs...1E.194P}
{Padovani} P.,  2017, \mn@doi [Nature Astronomy] {10.1038/s41550-017-0194},
  \href {http://adsabs.harvard.edu/abs/2017NatAs...1E.194P} {1, 0194}

\bibitem[\protect\citeauthoryear{Paraficz et~al.,}{Paraficz
  et~al.}{2018}]{paraficz2018alma}
Paraficz D.,  et~al., 2018, Astronomy \& Astrophysics, 613, A34

\bibitem[\protect\citeauthoryear{{Perley} \& {Butler}}{{Perley} \&
  {Butler}}{2017}]{2017ApJS..230....7P}
{Perley} R.~A.,  {Butler} B.~J.,  2017, \mn@doi [\apjs]
  {10.3847/1538-4365/aa6df9}, \href
  {https://ui.adsabs.harvard.edu/abs/2017ApJS..230....7P} {230, 7}

\bibitem[\protect\citeauthoryear{{Planck Collaboration} et~al.,}{{Planck
  Collaboration} et~al.}{2016}]{2016A&A...594A..13P}
{Planck Collaboration} et~al., 2016, \mn@doi [\aap]
  {10.1051/0004-6361/201525830}, \href
  {https://ui.adsabs.harvard.edu/abs/2016A&A...594A..13P} {594, A13}

\bibitem[\protect\citeauthoryear{Pooley, Blackburne, Rappaport, Schechter  \&
  fai Fong}{Pooley et~al.}{2006}]{0004-637X-648-1-67}
Pooley D.,  Blackburne J.~A.,  Rappaport S.,  Schechter P.~L.,   fai Fong W.,
  2006, The Astrophysical Journal, 648, 67

\bibitem[\protect\citeauthoryear{Popovi\'c, Afanasiev, Moiseev, Smirnova,
  Simi\'c, Savi\'c, Mediavilla  \& Fian}{Popovi\'c et~al.}{2020}]{Popovi__2020}
Popovi\'c L.~v.,  Afanasiev V.~L.,  Moiseev A.,  Smirnova A.,  Simi\'c S.,
  Savi\'c D.,  Mediavilla E.~G.,   Fian C.,  2020, \mn@doi [Astronomy &
  Astrophysics] {10.1051/0004-6361/201936088}, 634, A27

\bibitem[\protect\citeauthoryear{{Quinn} et~al.,}{{Quinn}
  et~al.}{2016}]{2016MNRAS.459.2394Q}
{Quinn} J.,  et~al., 2016, \mn@doi [\mnras] {10.1093/mnras/stw773}, \href
  {https://ui.adsabs.harvard.edu/\#abs/2016MNRAS.459.2394Q} {459, 2394}

\bibitem[\protect\citeauthoryear{{Rankine}, {Matthews}, {Hewett}, {Banerji},
  {Morabito}  \& {Richards}}{{Rankine} et~al.}{2021}]{2021MNRAS.502.4154R}
{Rankine} A.~L.,  {Matthews} J.~H.,  {Hewett} P.~C.,  {Banerji} M.,  {Morabito}
  L.~K.,   {Richards} G.~T.,  2021, \mn@doi [\mnras] {10.1093/mnras/stab302},
  \href {https://ui.adsabs.harvard.edu/abs/2021MNRAS.502.4154R} {502, 4154}

\bibitem[\protect\citeauthoryear{Rela{\~{n}}o \& Kennicutt}{Rela{\~{n}}o \&
  Kennicutt}{2009}]{Rela_o_2009}
Rela{\~{n}}o M.,  Kennicutt R.~C.,  2009, \mn@doi [The Astrophysical Journal]
  {10.1088/0004-637x/699/2/1125}, 699, 1125

\bibitem[\protect\citeauthoryear{{Richards}, {McCaffrey}, {Kimball}, {Rankine},
  {Matthews}, {Hewett}  \& {Rivera}}{{Richards}
  et~al.}{2021}]{2021arXiv210607783R}
{Richards} G.~T.,  {McCaffrey} T.~V.,  {Kimball} A.,  {Rankine} A.~L.,
  {Matthews} J.~H.,  {Hewett} P.~C.,   {Rivera} A.~B.,  2021, arXiv e-prints,
  \href {https://ui.adsabs.harvard.edu/abs/2021arXiv210607783R} {p.
  arXiv:2106.07783}

\bibitem[\protect\citeauthoryear{{Rodr{\'\i}guez-Ardila} \&
  {Mazzalay}}{{Rodr{\'\i}guez-Ardila} \&
  {Mazzalay}}{2006}]{2006MNRAS.367L..57R}
{Rodr{\'\i}guez-Ardila} A.,  {Mazzalay} X.,  2006, \mn@doi [\mnras]
  {10.1111/j.1745-3933.2006.00139.x}, \href
  {https://ui.adsabs.harvard.edu/abs/2006MNRAS.367L..57R} {367, L57}

\bibitem[\protect\citeauthoryear{{Rupen}, {van Gorkom}, {Knapp}, {Gunn}  \&
  {Schneider}}{{Rupen} et~al.}{1987}]{1987AJ.....94...61R}
{Rupen} M.~P.,  {van Gorkom} J.~H.,  {Knapp} G.~R.,  {Gunn} J.~E.,
  {Schneider} D.~P.,  1987, \mn@doi [\aj] {10.1086/114447}, \href
  {http://adsabs.harvard.edu/abs/1987AJ.....94...61R} {94, 61}

\bibitem[\protect\citeauthoryear{Rusin \& Ma}{Rusin \&
  Ma}{2001}]{rusin2001constraints}
Rusin D.,  Ma C.-P.,  2001, The Astrophysical Journal Letters, 549, L33

\bibitem[\protect\citeauthoryear{{Sabater, J.} et~al.,}{{Sabater, J.}
  et~al.}{2019}]{lotssagn}
{Sabater, J.} et~al., 2019, \mn@doi [A\&A] {10.1051/0004-6361/201833883}, 622,
  A17

\bibitem[\protect\citeauthoryear{{Sandage}}{{Sandage}}{1965}]{1965ApJ...141.1560S}
{Sandage} A.,  1965, \mn@doi [\apj] {10.1086/148245}, \href
  {https://ui.adsabs.harvard.edu/abs/1965ApJ...141.1560S} {141, 1560}

\bibitem[\protect\citeauthoryear{{Schechter} et~al.,}{{Schechter}
  et~al.}{1997}]{1997ApJ...475L..85S}
{Schechter} P.~L.,  et~al., 1997, \mn@doi [\apjl] {10.1086/310478}, \href
  {http://adsabs.harvard.edu/abs/1997ApJ...475L..85S} {475, L85}

\bibitem[\protect\citeauthoryear{{Schmidt}}{{Schmidt}}{1963}]{1963Natur.197.1040S}
{Schmidt} M.,  1963, \mn@doi [\nat] {10.1038/1971040a0}, \href
  {http://adsabs.harvard.edu/abs/1963Natur.197.1040S} {197, 1040}

\bibitem[\protect\citeauthoryear{{Serjeant} \& {Hatziminaoglou}}{{Serjeant} \&
  {Hatziminaoglou}}{2009}]{2009MNRAS.397..265S}
{Serjeant} S.,  {Hatziminaoglou} E.,  2009, \mn@doi [\mnras]
  {10.1111/j.1365-2966.2009.14431.x}, \href
  {https://ui.adsabs.harvard.edu/abs/2009MNRAS.397..265S} {397, 265}

\bibitem[\protect\citeauthoryear{Sharon et~al.,}{Sharon
  et~al.}{2005}]{sharon2005discovery}
Sharon K.,  et~al., 2005, The Astrophysical Journal Letters, 629, L73

\bibitem[\protect\citeauthoryear{{Shimwell} et~al.,}{{Shimwell}
  et~al.}{2019}]{2019A&A...622A...1S}
{Shimwell} T.~W.,  et~al., 2019, \mn@doi [\aap] {10.1051/0004-6361/201833559},
  \href {https://ui.adsabs.harvard.edu/abs/2019A&A...622A...1S} {622, A1}

\bibitem[\protect\citeauthoryear{{Sluse}, {Chantry, V.}, {Magain, P.},
  {Courbin, F.}  \& {Meylan, G.}}{{Sluse} et~al.}{2012}]{sluse1115}
{Sluse} D.,  {Chantry, V.} {Magain, P.} {Courbin, F.}  {Meylan, G.} 2012,
  \mn@doi [A\&A] {10.1051/0004-6361/201015844}, 538, A99

\bibitem[\protect\citeauthoryear{Sonnenfeld, Treu, Gavazzi, Marshall, Auger,
  Suyu, Koopmans  \& Bolton}{Sonnenfeld et~al.}{2012}]{0004-637X-752-2-163}
Sonnenfeld A.,  Treu T.,  Gavazzi R.,  Marshall P.~J.,  Auger M.~W.,  Suyu
  S.~H.,  Koopmans L. V.~E.,   Bolton A.~S.,  2012, The Astrophysical Journal,
  752, 163

\bibitem[\protect\citeauthoryear{Sonnenfeld, Treu, Gavazzi, Suyu, Marshall,
  Auger  \& Nipoti}{Sonnenfeld et~al.}{2013}]{0004-637X-777-2-98}
Sonnenfeld A.,  Treu T.,  Gavazzi R.,  Suyu S.~H.,  Marshall P.~J.,  Auger
  M.~W.,   Nipoti C.,  2013, The Astrophysical Journal, 777, 98

\bibitem[\protect\citeauthoryear{Sonnenfeld, Treu, Marshall, Suyu, Gavazzi,
  Auger  \& Nipoti}{Sonnenfeld et~al.}{2015}]{0004-637X-800-2-94}
Sonnenfeld A.,  Treu T.,  Marshall P.~J.,  Suyu S.~H.,  Gavazzi R.,  Auger
  M.~W.,   Nipoti C.,  2015, The Astrophysical Journal, 800, 94

\bibitem[\protect\citeauthoryear{Sopp \& Alexander}{Sopp \&
  Alexander}{1991}]{doi:10.1093/mnras/251.1.14P}
Sopp H.~M.,  Alexander P.,  1991, \mn@doi [Monthly Notices of the Royal
  Astronomical Society] {10.1093/mnras/251.1.14P}, 251, 14P

\bibitem[\protect\citeauthoryear{{Spilker} et~al.,}{{Spilker}
  et~al.}{2016}]{spilker16a}
{Spilker} J.~S.,  et~al., 2016, \mn@doi [\apj] {10.3847/0004-637X/826/2/112},
  \href {http://adsabs.harvard.edu/abs/2016ApJ...826..112S} {826, 112}

\bibitem[\protect\citeauthoryear{Spingola, McKean, Deller  \& Moldon}{Spingola
  et~al.}{2019}]{spingola2019gravitational}
Spingola C.,  McKean J.~P.,  Deller A.,   Moldon J.,  2019, Gravitational
  lensing at milliarcsecond angular resolution with VLBI observations
  (\mn@eprint {arXiv} {1902.07046})

\bibitem[\protect\citeauthoryear{{Stacey} et~al.,}{{Stacey}
  et~al.}{2018a}]{2018MNRAS.476.5075S}
{Stacey} H.~R.,  et~al., 2018a, \mn@doi [\mnras] {10.1093/mnras/sty458}, \href
  {http://adsabs.harvard.edu/abs/2018MNRAS.476.5075S} {476, 5075}

\bibitem[\protect\citeauthoryear{Stacey et~al.,}{Stacey
  et~al.}{2018b}]{ae7b5bac19b64fffa43b0dcb87037cbf}
Stacey H.,  et~al., 2018b, \mn@doi [Monthly Notices of the Royal Astronomical
  Society] {10.1093/mnras/sty458}, 476, 5075

\bibitem[\protect\citeauthoryear{{Stacey} et~al.,}{{Stacey}
  et~al.}{2019}]{hannah2}
{Stacey} H.~R.,  et~al., 2019, \mn@doi [A\&A] {10.1051/0004-6361/201833967},
  622, A18

\bibitem[\protect\citeauthoryear{Stacey et~al.,}{Stacey
  et~al.}{2020}]{10.1093/mnras/staa3433}
Stacey H.~R.,  et~al., 2020, \mn@doi [Monthly Notices of the Royal Astronomical
  Society] {10.1093/mnras/staa3433}, 500, 3667

\bibitem[\protect\citeauthoryear{{Suyu} et~al.,}{{Suyu}
  et~al.}{2014}]{2014ApJ...788L..35S}
{Suyu} S.~H.,  et~al., 2014, \mn@doi [\apjl] {10.1088/2041-8205/788/2/L35},
  \href {https://ui.adsabs.harvard.edu/abs/2014ApJ...788L..35S} {788, L35}

\bibitem[\protect\citeauthoryear{Symeonidis, Giblin, Page, Pearson, Bendo,
  Seymour  \& Oliver}{Symeonidis et~al.}{2016}]{10.1093/mnras/stw667}
Symeonidis M.,  Giblin B.~M.,  Page M.~J.,  Pearson C.,  Bendo G.,  Seymour N.,
    Oliver S.~J.,  2016, \mn@doi [Monthly Notices of the Royal Astronomical
  Society] {10.1093/mnras/stw667}, 459, 257

\bibitem[\protect\citeauthoryear{{Tabatabaei} \& {Berkhuijsen}}{{Tabatabaei} \&
  {Berkhuijsen}}{2010}]{Tabatabaei19}
{Tabatabaei} F.~S.,  {Berkhuijsen} E.~M.,  2010, \mn@doi [A\&A]
  {10.1051/0004-6361/200913593}, 517, A77

\bibitem[\protect\citeauthoryear{Taylor, Dunlop, Hughes  \& Robson}{Taylor
  et~al.}{1996}]{10.1093/mnras/283.3.930}
Taylor G.~L.,  Dunlop J.~S.,  Hughes D.~H.,   Robson E.~I.,  1996, \mn@doi
  [Monthly Notices of the Royal Astronomical Society]
  {10.1093/mnras/283.3.930}, 283, 930

\bibitem[\protect\citeauthoryear{{Treu} \& {Koopmans}}{{Treu} \&
  {Koopmans}}{2002}]{2002ApJ...575...87T}
{Treu} T.,  {Koopmans} L.~V.~E.,  2002, \mn@doi [\apj] {10.1086/341216}, \href
  {http://cdsads.u-strasbg.fr/abs/2002ApJ...575...87T} {575, 87}

\bibitem[\protect\citeauthoryear{{Treu} \& {Koopmans}}{{Treu} \&
  {Koopmans}}{2003}]{2003MNRAS.343L..29T}
{Treu} T.,  {Koopmans} L.~V.~E.,  2003, \mn@doi [\mnras]
  {10.1046/j.1365-8711.2003.06858.x}, \href
  {http://cdsads.u-strasbg.fr/abs/2003MNRAS.343L..29T} {343, L29}

\bibitem[\protect\citeauthoryear{{Treu} \& {Koopmans}}{{Treu} \&
  {Koopmans}}{2004}]{2004ApJ...611..739T}
{Treu} T.,  {Koopmans} L.~V.~E.,  2004, \mn@doi [\apj] {10.1086/422245}, \href
  {http://cdsads.u-strasbg.fr/abs/2004ApJ...611..739T} {611, 739}

\bibitem[\protect\citeauthoryear{{Tsvetkova} et~al.,}{{Tsvetkova}
  et~al.}{2010}]{2010MNRAS.406.2764T}
{Tsvetkova} V.~S.,  et~al., 2010, \mn@doi [\mnras]
  {10.1111/j.1365-2966.2010.16882.x}, \href
  {https://ui.adsabs.harvard.edu/#abs/2010MNRAS.406.2764T} {406, 2764}

\bibitem[\protect\citeauthoryear{Utomo, Chiang, Leroy, Sandstrom  \&
  Chastenet}{Utomo et~al.}{2019}]{Utomo_2019}
Utomo D.,  Chiang I.-D.,  Leroy A.~K.,  Sandstrom K.~M.,   Chastenet J.,  2019,
  \mn@doi [The Astrophysical Journal] {10.3847/1538-4357/ab05d3}, 874, 141

\bibitem[\protect\citeauthoryear{{Vegetti}, {Lagattuta}, {McKean}, {Auger},
  {Fassnacht}  \& {Koopmans}}{{Vegetti} et~al.}{2012}]{2012Natur.481..341V}
{Vegetti} S.,  {Lagattuta} D.~J.,  {McKean} J.~P.,  {Auger} M.~W.,  {Fassnacht}
  C.~D.,   {Koopmans} L.~V.~E.,  2012, \mn@doi [\nat] {10.1038/nature10669},
  \href {http://adsabs.harvard.edu/abs/2012Natur.481..341V} {481, 341}

\bibitem[\protect\citeauthoryear{Villar-Martin, de Young, Alonso-Herrero, Allen
   \& Binette}{Villar-Martin et~al.}{2001}]{10.1046/j.1365-8711.2001.04916.x}
Villar-Martin M.,  de Young D.,  Alonso-Herrero A.,  Allen M.,   Binette L.,
  2001, \mn@doi [Monthly Notices of the Royal Astronomical Society]
  {10.1046/j.1365-8711.2001.04916.x}, 328, 848

\bibitem[\protect\citeauthoryear{Villar-Mart\'{i}n et~al.,}{Villar-Mart\'{i}n
  et~al.}{2017}]{10.1093/mnras/stx2209}
Villar-Mart\'{i}n M.,  et~al., 2017, \mn@doi [Monthly Notices of the Royal
  Astronomical Society] {10.1093/mnras/stx2209}, 472, 4659

\bibitem[\protect\citeauthoryear{Wallington \& Narayan}{Wallington \&
  Narayan}{1993}]{wallington1993influence}
Wallington S.,  Narayan R.,  1993, The Astrophysical Journal, 403, 517

\bibitem[\protect\citeauthoryear{{Weymann}, {Latham}, {Angel}, {Green},
  {Liebert}, {Turnshek}, {Turnshek}  \& {Tyson}}{{Weymann}
  et~al.}{1980}]{1980Natur.285..641W}
{Weymann} R.~J.,  {Latham} D.,  {Angel} J.~R.~P.,  {Green} R.~F.,  {Liebert}
  J.~W.,  {Turnshek} D.~A.,  {Turnshek} D.~E.,   {Tyson} J.~A.,  1980, \mn@doi
  [\nat] {10.1038/285641a0}, \href
  {http://adsabs.harvard.edu/abs/1980Natur.285..641W} {285, 641}

\bibitem[\protect\citeauthoryear{{White}, {Jarvis}, {Kalfountzou},
  {Hardcastle}, {Verma}, {Cao Orjales}  \& {Stevens}}{{White}
  et~al.}{2017}]{2017MNRAS.468..217W}
{White} S.~V.,  {Jarvis} M.~J.,  {Kalfountzou} E.,  {Hardcastle} M.~J.,
  {Verma} A.,  {Cao Orjales} J.~M.,   {Stevens} J.,  2017, \mn@doi [\mnras]
  {10.1093/mnras/stx284}, \href
  {http://adsabs.harvard.edu/abs/2017MNRAS.468..217W} {468, 217}

\bibitem[\protect\citeauthoryear{Williams \& Saha}{Williams \&
  Saha}{2004}]{williams2004models}
Williams L.~L.,  Saha P.,  2004, The Astronomical Journal, 128, 2631

\bibitem[\protect\citeauthoryear{{Winn}, {Rusin}  \& {Kochanek}}{{Winn}
  et~al.}{2004}]{2004Natur.427..613W}
{Winn} J.~N.,  {Rusin} D.,   {Kochanek} C.~S.,  2004, \mn@doi [\nat]
  {10.1038/nature02279}, \href
  {https://ui.adsabs.harvard.edu/\#abs/2004Natur.427..613W} {427, 613}

\bibitem[\protect\citeauthoryear{{Wucknitz} \& {Volino}}{{Wucknitz} \&
  {Volino}}{2008}]{2008arXiv0811.3421W}
{Wucknitz} O.,  {Volino} F.,  2008, preprint, \href
  {http://adsabs.harvard.edu/abs/2008arXiv0811.3421W} {} (\mn@eprint {arXiv}
  {0811.3421})

\bibitem[\protect\citeauthoryear{Wyithe, Agol  \& Fluke}{Wyithe
  et~al.}{2002}]{10.1046/j.1365-8711.2002.05252.x}
Wyithe J. S.~B.,  Agol E.,   Fluke C.~J.,  2002, \mn@doi [Monthly Notices of
  the Royal Astronomical Society] {10.1046/j.1365-8711.2002.05252.x}, 331, 1041

\bibitem[\protect\citeauthoryear{{Xu}, {Sun}  \& {Xue}}{{Xu}
  et~al.}{2020}]{2020ApJ...894...21X}
{Xu} J.,  {Sun} M.,   {Xue} Y.,  2020, \mn@doi [\apj]
  {10.3847/1538-4357/ab811a}, \href
  {https://ui.adsabs.harvard.edu/abs/2020ApJ...894...21X} {894, 21}

\bibitem[\protect\citeauthoryear{{Zakamska} \& {Greene}}{{Zakamska} \&
  {Greene}}{2014}]{2014MNRAS.442..784Z}
{Zakamska} N.~L.,  {Greene} J.~E.,  2014, \mn@doi [\mnras]
  {10.1093/mnras/stu842}, \href
  {http://adsabs.harvard.edu/abs/2014MNRAS.442..784Z} {442, 784}

\bibitem[\protect\citeauthoryear{{Zakamska} et~al.,}{{Zakamska}
  et~al.}{2016}]{2016MNRAS.455.4191Z}
{Zakamska} N.~L.,  et~al., 2016, \mn@doi [\mnras] {10.1093/mnras/stv2571},
  \href {https://ui.adsabs.harvard.edu/abs/2016MNRAS.455.4191Z} {455, 4191}

\bibitem[\protect\citeauthoryear{{de Jong}, {Klein}, {Wielebinski}  \&
  {Wunderlich}}{{de Jong} et~al.}{1985}]{1985A&A...147L...6D}
{de Jong} T.,  {Klein} U.,  {Wielebinski} R.,   {Wunderlich} E.,  1985, \aap,
  \href {https://ui.adsabs.harvard.edu/abs/1985A&A...147L...6D} {147, L6}

\makeatother
\end{thebibliography}






\bsp	
\label{lastpage}
\end{document}